\documentclass[aps,floatfix,superscriptaddress,reprint,10pt,prb]{revtex4-1}
\usepackage{bm}
\usepackage{amsmath}
\usepackage{amssymb}
\usepackage{multirow}
\usepackage{empheq}
\usepackage{graphicx}
\usepackage{mathrsfs}
\usepackage{amsfonts}
\usepackage{amsthm}
\usepackage{color}
\usepackage{bigints}
\usepackage{txfonts}
\usepackage{hyperref}
\hypersetup{
	unicode=false,          % non-Latin characters in Acrobat?s bookmarks
	pdftoolbar=true,        % show Acrobat?s toolbar?
	pdfmenubar=true,        % show Acrobat?s menu?
	pdffitwindow=false,     % window fit to page when opened
	pdfstartview={FitH},    % fits the width of the page to the window
	pdftitle={My title},    % title
	pdfauthor={Author},     % author
	pdfsubject={Subject},   % subject of the document
	pdfcreator={Creator},   % creator of the document
	pdfproducer={Producer}, % producer of the document
	pdfkeywords={keyword1} {key2} {key3}, % list of keywords
	pdfnewwindow=true,      % links in new PDF window
	colorlinks=false,       % false: boxed links; true: colored links
	linkcolor=red,          % color of internal links (change box color with linkbordercolor)
	citecolor=green,        % color of links to bibliography
	filecolor=magenta,      % color of file links
	urlcolor=cyan           % color of external links
}
\makeatletter \tolerance = 10000 \tolerance = 10000
\usepackage{color}

\begin{document}
	
	\title{On the sample-dependent minimal conductivity in weakly disordered graphene}
	
	\author{Weiwei Chen}
	\affiliation{Institute of Natural Sciences, Westlake Institute for Advanced Study, 18 Shilongshan Road, Hangzhou 310024, China}
	\affiliation{School of Science, Westlake University, 18 Shilongshan Road, Hangzhou 310024,  China}

	\author{Yedi Shen}
	\affiliation{Hefei National Laboratory for Physical Sciences at the Microscale, and Synergetic Innovation Center of Quantum Information and Quantum Physics, University of Science and Technology of China, Hefei, Anhui 230026, China}
	
	\author{Bo Fu}
	\affiliation{Department of Physics, The University of Hong Kong, Pokfulam Road, Hong Kong, China}
	
	\author{Qinwei Shi}
	\thanks{E-mail: phsqw@ustc.edu.cn}
	\affiliation{Hefei National Laboratory for Physical Sciences at the Microscale, and Synergetic Innovation Center of Quantum Information and Quantum Physics, University of Science and Technology of China, Hefei, Anhui 230026, China}
	
	\author{W. Zhu}
	\thanks{E-mail: zhuwei@westlake.edu.cn}
	\affiliation{ Key Laboratory for Quantum Materials of Zhejiang Province, Department of Physics, School of Science, Westlake University, Hangzhou 310024, China}
	%\affiliation{Institute of Natural Sciences, Westlake Institute for Advanced Study, 18 Shilongshan Road, Hangzhou 310024,  China}

	\begin{abstract}
		We present a unified understanding of the experimentally observed minimal dc conductivity in weakly disordered graphene.
		Firstly, based on linear response theory, we reveal that randomness or disorder inevitably induces momentum  dependent corrections to the electron self-energy function, which naturally yields a sample-dependent minimal conductivity. 
		Taking the long-ranged Gaussian and Coulomb potentials as examples, we derive the momentum dependent self-energy function within the Born approximation, and further validate it via numerical simulations using the large-scale Lanczos algorithm. The explicit momentum dependences of the self-energy on the intensity, concentration and range of potential are critically addressed. Therefore, our results provide a reasonable interpretation of the sample-dependent minimal conductivity observed in graphene samples.
	\end{abstract}

	\maketitle

	\clearpage
	\noindent\textbf{Introduction}\\
	The earliest experimental measurements on graphene reported a finite dc conductivity $\sigma_{\text{min}}\approx \frac{4e^2}{h}$ at the Dirac point  at low temperature \cite{Novoselov,Zhang2005,Novoselov2005}, dubbed as the minimal conductivity. The observed value is larger than the earlier prediction of $\sigma_{\text{min}}=\frac{4e^2}{\pi h}$ \cite{Shon,Ludwig,Nersesyan}, giving rise to the famous ``missing $\pi$" problem \cite{Sarma2011}. To address this problem, subsequent experiments have been carried out \cite{Bolotin,Du,Zhang2005,Morozov,Dean,Zomer,Mayorov2012,Nuno,WangLujun,Tan}, which show that the minimal conductivity is strongly sample-dependent, i.e. $\sigma_{\text{min}}=C\frac{4e^2}{\pi h}$ with the factor $C$ varying from 1.7 to 10, indicating the crucial role of randomness or disorder.

	%However, more experiments show that the minimum conductivity can exhibits value in the range of (2-15)$\frac {e^2}{h}$ and clearly display the sample-dependence behavior.  More interestingly, the similar behavior has been reported in a number of transport measurements in ultrahigh-mobility samples, although the key difference with early measurements is that the minimum conductivity shows strong temperature dependence. The minimum conductivity in ultrahigh-mobility samples at low temperature can be described as $\sigma_{\text{min}}=C\frac{4e^2}{\pi h}$. The factor $C$ varies from 1.6 to 10 reported by different research group \cite{Bolotin,Du,Zhang2005,Morozov,Dean,Zomer,Mayorov2012,Nuno,WangLujun,Tan}, indicating that the impurity scattering should play a crucial roles to determining the minimum conductivity.
	
	The transport properties of two-dimensional disordered Dirac fermions have been intensively studied for the d-wave superconductivity in the cuprate superconductors \cite{Patrick,Durst} and the plateau transition in the integer quantum Hall effect \cite{Ludwig}. 
	The  discovery of graphene \cite{Novoselov,Zhang2005,Novoselov2005} rejuvenated this problem.
	%Then, experimental discovery of graphene \cite{Novoselov,Novoselov2005} not only provides an ideal platform to measure the minimal conductivity  because it is easy to control the carriers density to access the Dirac point, but also stimulates intensive theoretical exploration of the Dirac point physics. 
	The previous theoretical results for $\sigma_{\text{min}}$ can be mainly summarized as:
	(i) a scattering-independent value $\sigma_{\text{min}}=\frac{4e^2}{\pi h}$ \cite{Peres2006,Shon,Ludwig,Katsnelson2006,Nersesyan,Tworzydlo,Ostrovsky,Fedorenko}, which is independent of the strength and nature of the disorder;
	(ii) a universal value $\sigma_{\text{min}}=\frac{4e^2}{ h}$, which is due to the quantum criticality of graphene in the vicinity of the Dirac point \cite{Ostrovsky2007,Ostrovsky2007prl,Schuessler};
	(iii) a universal value $\sigma_{\text{min}}=\frac{\pi e^2}{2h}$ \cite{Ziegler,Ryu}, from the ac Kubo formula, derived by removing the smearing of the single-particle Green's functions before taking the dc limit;
	(iv) a disorder-dependent $\sigma_{\text{min}}$ \cite{Shon,Trushin,Adam2009,Sarma2011,Nomura,Noro,Radchenko,Bardarson,Rycerz}.
	In particular, although some previous studies have yielded a disordered-dependent minimal conductivity, most of these are based on semi-classical Boltzmann transport theory \cite{Shon,Trushin,Adam2009,Sarma2011}, ignoring the disorder induced quantum corrections near the Dirac point. 
	%Besides, it is found that a long-range randomness leads to emergence of a topological term in the corresponding field theory, as a consequence, the system is at a quantum critical point with a universal value of the conductivity of the order of $e^2/h$  \cite{Ostrovsky2007,Ostrovsky2007prl,Schuessler}. 
	%%\oim{Check the references: Why do Ref. 21-23 include in the disorder-dependent $\sigma_{min}$ above?   Or, shall we add (iii) with $\sigma_{min}=4e^2/h$ ? }
	Since Boltzmann theory is not applicable around the Dirac point \cite{Sarma2011}, a fully quantum mechanical treatment is highly desirable in order to address the minimal conductivity.

	In parallel, numerical calculations have also addressed this problem using various approaches \cite{Nomura,Noro,Radchenko,Bardarson,Rycerz}.
	Nomura and MacDonald numerically calculated $\sigma_{\text{min}}$ using the Kubo formula and found that $\sigma_{\text{min}}$ is a few times larger for long-range Coulomb scatterers than for short-range scatterers \cite{Nomura}. Similar results were also obtained by numerically calculating the transmission matrix \cite{Bardarson,Rycerz}. Noro \textit{et al.} find that the minimal conductivity at the Dirac point remains universal in the clean limit, but increases with disorder and becomes non-universal for long-range scatterers, by numerically solving the self-energy and current vertex function \cite{Noro}. 
	These numerical results, which are beyond the aforementioned theoretical descriptions, urgently call for a reasonable analytical interpretation of the minimal conductivity.

	In this paper, we offer a general picture for understanding the sample-dependent minimal conductivity by studying the effect of long-ranged random potentials. The long-ranged randomness  could be realized by screened charges in the substrate \cite{Rycerz,Fan,Chen2008}, local strain fluctuations \cite{Nuno,WangLujun}, and other defects that vary smoothly on the atomic scale \cite{Tan,Klos}. Recently, it has been confirmed that random strain fluctuations are the dominant source of disorder for high-quality graphene on many different substrates, from spatially resolved Raman spectroscopy measurements, where the out-of-plane corrugation (stain) could induce a long-ranged scalar potential \cite{Nuno,WangLujun}. We mainly consider the long-ranged Gaussian potential, which is a simple model of atomic-scale random fluctuations, and the Gaussian smoothing is chosen for computational convenience. We also checked that the results are robust against the form of the random potentials by considering the long-range Coulomb potential (shown in Supplemental Material Sec.~S2~C). In the presence of long-ranged random potentials, we elucidate that the self-energy is momentum-dependent, which is overlooked in the existing literature. Crucially, the electric minimal conductivity, evaluated via the standard Kubo formula under disorder configuration average by taking into account the corrected self-energy, obeys the relationship $\sigma_{\text{min}}=\frac{1}{(1-\alpha)^2}\frac{4e^2}{\pi h}$ [Eq.(11)], where the leading momentum-dependent correction is parameterized by a dimensionless parameter $\alpha= K_0\frac{\xi^2k_c^2}{8\pi}$ [Eq.(19)], which depends on the disorder strength and the spatial range of the potential (see below for details). This finding, in sharp contrast to the disorder-independent values in previous studies \cite{Nersesyan,Ludwig,Shon,Peres2006,Ostrovsky,Katsnelson2006,Tworzydlo,Fedorenko,Ziegler}, provides a simple and physically appealing explanation for the observed minimum conductivity in experiments.

	~\\
	\noindent\textbf{Minimal conductivity with momentum dependent self-energy}
	
	The charge carriers of graphene near half filling can be modeled by a Dirac Hamiltonian
	\begin{equation}
		H=\hbar v_f\bm{\sigma}\cdot\bm{k}+U(\bm{r}),
		\label{Hamiltonian0}
	\end{equation}
	where $v_f\approx10^6$ $m\ s^{-1}$ is the Fermi velocity, $\bm{\sigma}=(\sigma_x,\sigma_y)$ are the Pauli matrices of pseudospin (sublattice), $\bm{k}=(k_x,k_y)$ is a two-component wave vector,
	and $U(\bm{r})$ describes the long-ranged random potential that is experienced by the Dirac electrons. In the pure case,
	the eigenvalues and eigenstates of the Hamiltonian Eq.~(\ref{Hamiltonian0}) are
	\begin{equation}
		E_{\bm{k}s}=s\hbar v_f k;\ \ \ 	\Psi_{\bm{k}s}(\bm{r})=\langle \bm{r}|\bm{k}s\rangle=
		\frac{e^{i\bm{k}\cdot\bm{r}}}{\sqrt{2\mathcal{V}}}\left(\begin{array}{c}
			1\\se^{i\theta_{\bm{k}}}
		\end{array}
		\right)
	\end{equation}
	where $s=\pm$ denotes the band index ($s=+1$ for the conduction band and $s=-1$ for the valence band), $\mathcal{V}$ is the sample area, and $\theta_{\bm{k}}=\arctan(k_y/k_x)$ is the angle of the wave vector $\bm{k}$ down from the positive $x$-axis. The velocity $\frac{1}{\hbar}\frac{\partial H}{\partial k_x}$ along the x direction in the eigenstates basis is
	\begin{equation}\label{vx}
		v_x(\bm{k})=v_f(\cos\theta_{\bm{k}}\sigma_z+\sin\theta_{\bm{k}}\sigma_y).
	\end{equation}
	
	For simplicity, we model the random potential by  $U(\bm{r})=\sum_{i=1}^{N_{\text{imp}}}\pm u_0e^{-|\bm{r}-\bm{R}_i|^2/\xi^2}$, where $\bm{R}_i$ and $N_{\text{imp}}$ are the location and total number of impurities, $\xi$ is the characteristic length scale and the strength $\pm u_0$ is randomly distributed with equal probability. We also checked that the results shown below are not sensitive to the form of the potentials. To do so, we checked the long-range screened Coulomb potential and obtained similar results (shown in Supplemental Material Sec.~S2~C).
	
	In general, the electron self-energy induced by the impurity scattering is both momentum- and energy-dependent.  For the short-ranged disorder potential, the momentum dependence of the self-energy is negligible \cite{Shon,Ludwig,Nersesyan,Sarma2011}. For the long-range disorder potential, the self-energy is usually treated in the on-shell approximation ($E=E_{\bm{k}s}$) \cite{Adam}, i.e. the momentum dependence of the self-energy is replaced by the single-particle energy in the clean limit. 
	One of motivations for this paper is to elucidate that this on-shell approximation is not well justified around the Dirac point, leading to a momentum-dependent self-energy function (We present both analytical and numerical evidence to support this in the sections below). Here, we first reveal the influence of the momentum-dependent self-energy function on the minimal conductivity.
	%However, the numerical calculations that we will present below show that  self-energy near the Dirac point is indeed dependent on the momentum. 
	Accordingly, for the discussion, we can assume that the self-energy function contains a nonzero momentum-dependent contribution,
	\begin{equation}\label{re-k-E}
		\Sigma(\bm{k}s,E)\approx \Sigma_1(E)-\alpha s\hbar v_fk+i\Sigma_2(E)
	\end{equation}
	where $s=\pm$ is the band index, while $\Sigma_1(E)$ and $\Sigma_2(E)$ describe the real and imaginary parts of the energy-dependent terms. 
	%More importantly, we introduce a leading order term proportional to the magnitude of the wave vector in the real part of self-energy to reflect the momentum dependence of self-energy function. 
	Since the long-ranged random scalar disorder we considered here does not on average break the particle-hole symmetry \cite{Altland1997,Ostrovsky,Fradkin,Aleiner}, the self-energy in the eigenstate basis satisfies the relations, ${\rm Re}\Sigma(\bm{k}+,E)=-{\rm Re}\Sigma(\bm{k}-,-E)$ and ${\rm Im}\Sigma(\bm{k}+,E)={\rm Im}\Sigma(\bm{k}-,-E)$, which preserves the dispersion relation and broadens the width of the conduction and valence bands symmetrically about the Dirac point \cite{Hu}.
	
	Now we study the influence of this momentum dependent self-energy function on the minimal conductivity. Based on linear response theory, we can calculate the longitudinal conductivity at zero temperature using the Kubo formula \cite{Bruus} (see Supplemental Material Sec.~S1)
	\begin{equation}\begin{aligned}\label{Gaussian-dc-sigma-total}
			\sigma_{xx}(E)=\sigma_{xx}^{RA}(E)-{\rm Re}\left[\sigma_{xx}^{RR}(E)\right]
	\end{aligned}\end{equation}
	with
	\begin{eqnarray}
		\label{Gaussian-dc-sigma-RA}
		\sigma_{xx}^{RA}(E)&=&4\frac{e^2\hbar}{2\pi\mathcal{V}}{\rm Tr}\left[G^R(E)v_xG^A(E)v_x\right]_c\\
		\label{Gaussian-dc-sigma-RR}
		\sigma_{xx}^{RR}(E)&=&4\frac{e^2\hbar}{2\pi\mathcal{V}}{\rm Tr}\left[G^R(E)v_xG^R(E)v_x\right]_c
	\end{eqnarray}
	where the factor 4 denotes the degeneracy of the real spin and valley, ${\rm Tr}[\cdots]$ means the trace over both wave vector and pseudospin (sublattice) spaces, and the subscript $c$ indicates a disorder configuration average. Here the RA contribution is analogous to the transport of classical particles within the relaxation time approximation and gives results close to conventional Boltzmann theory, while the RR contribution comes from calculations with a fully quantum mechanism and can be ignored in the limit $E\tau\gg1$ (but not when $E\tau \ll 1$). $G^{R/A}(E)$ is the full retarded/advanced Green's function. In the eigenstate basis, it is
	\begin{equation}\label{green-G}
		G^{R/A}(\bm{k},E)=\left(\begin{array}{cc}
			g^{R/A}_{+}(\bm{k},E)&0\\0&g^{R/A}_{-}(\bm{k},E)
		\end{array}\right)
	\end{equation}
	where
	\begin{equation}\label{green-g}
		g^{R/A}_s(\bm{k},E)=\frac{1}{a-(1-\alpha)s\hbar v_fk\pm i\eta},\,\,\,\,\,
		%\end{equation}
		%and
		%\begin{equation}
		\left\{\begin{array}{l}
			a=E-\Sigma_1(E)\\
			\eta=-\Sigma_2(E)\\
		\end{array}\right.
	\end{equation}
	Plugging Eq.~(\ref{vx}), Eq.~(\ref{green-G}) and Eq.~(\ref{green-g}) into the Kubo formula Eq.~(\ref{Gaussian-dc-sigma-total}), we obtain the conductivity
	\begin{equation}
		\sigma_{xx}(E)=\frac{2e^2}{\pi h}\frac{1}{(1-\alpha)^2}(1+\frac{a}{\eta}\arctan\frac{a}{\eta})
		+\frac{2e^2}{\pi h}\frac{1}{(1-\alpha)^2}\frac{\eta}{a}\arctan\frac{a}{\eta}
	\end{equation}
	where the first and second terms denote the electron-hole incoherent and coherent contributions, respectively. At the Dirac point ($E=0$), these two parts contribute equally and the minimal conductivity is obtained as
	\begin{equation}\label{min-sigma}
		\sigma_{\text{min}}=\frac{1}{(1-\alpha)^2}\frac{4e^2}{\pi h}.
	\end{equation}
	
	Here we would like to give some remarks. First,  Eq.~(\ref{min-sigma}) contains both electron-hole coherent and incoherent contributions, which is beyond traditional Boltzmann transport theory where the electron-hole incoherent contribution from the retarded-retarded ($RR$) channel is usually discarded \cite{Trushin}.
	% It is also worth noting that, even the electron-hole incoherent contribution of $\sigma_{\text{min}}$ comes from the retarded-retarded ($RR$) channel, indicating the minimal conductivity completely exceeds the traditional Boltzmann transport theory.
	Second, the influence of the momentum-dependent self-energy on the vertex correction is much smaller than the level of the bubble diagram (see Supplemental Materials Sec. S5). This is consistent with previous works showing that the vertex correction is negligible for calculations of the minimal conductivity \cite{Durst,Shon}. 
	%A plausible explanation is that the forward and back scattering are exactly complementary in the electron-hole incoherent and coherent contributions for the isotropic disorder potential at the Dirac point, while the vertex correction enhances the back scattering and weakens the forward scattering. 
	Third, from the result Eq.~(\ref{min-sigma}), it is clear that the minimal conductivity $\sigma_{\text{min}}$ is independent of the energy-dependent part of the self-energy function, i.e. $\Sigma_1(E)$ and $\Sigma_2(E)$, if they satisfy the condition, $\frac{a}{\eta}\to0$ when $E\to0$, but $\sigma_{\text{min}}$ is significantly influenced by the momentum dependent term. This behavior explains why the majority of previous theoretical studies predicted $\sigma_{\text{min}}=\frac{4e^2}{\pi h}$, independent of the strength and nature of the disorder \cite{Ludwig,Nersesyan,Ziegler}. 
	%For instance, the pioneering works on this problem often employ a constant broadening approximation and naturally satisfy the limit condition \cite{Patrick,Ludwig,Nersesyan,McCann,Ziegler}. 
	Fourth, for long range disorder, previous calculations within the self-consistent Born approximation reported the momentum contribution to the self-energy function, however, the role of this term on the transport was unfortunately washed out in the widely-used on-shell approximation \cite{Ostrovsky}. 
	%Fifth, compared with the research reporting a disorder-dependent minimal conductivity, the disorder induced finite density of state at the Dirac point is ignored, which leads to a finite and  disorder-dependent $\frac{a}{\eta}$ term when $E\to0$ \cite{Trushin}.
	Therefore, we claim that the application of the on-shell approximation around the Dirac point is questionable. Finally, with all of these non-trivial improvements, we point out that Eq.~(\ref{min-sigma}) shows that the minimal conductivity is enhanced by disorder, in contrast to the results from Boltzmann theory \cite{Shon,Adam2009,Trushin}. 
	One interpretation is that the presence of potential fluctuations smooth on the scale of the graphene lattice spacing increases the conductivity through quantum interference effects \cite{Adam,Nomura,Noro,Radchenko,Bardarson,Rycerz}.

	\noindent\textbf{Estimation of $\alpha$ by the Born approximation}\\
	In order to  theoretically incorporate insights gained from the momentum dependent self-energy, we take the long-ranged Gaussian potential as an example, with the correlation function as
	\begin{equation}\label{correlation-function}
		%\mathcal{K}(\bm{r}-\bm{r}')=K_0\frac{(\hbar v_f)^2}{2\pi\xi^2}e^{-|\bm{r}-\bm{r}'|^2/2\xi^2}
		\mathcal{K}(\bm{r}-\bm{r}')=\langle U(\bm{r})U(\bm{r}')\rangle=K_0\frac{(\hbar v_f)^2}{2\pi\xi^2}e^{-|\bm{r}-\bm{r}'|^2/2\xi^2}
	\end{equation}
	where $\xi$ is the correlation length and $K_0$ is a dimensionless parameter that parametrizes its magnitude. Its Fourier transformation form is
	\begin{equation}\begin{aligned}\label{Vq}
			\mathcal{K}(\bm{q})=& %\langle U(\bm{q})U(-\bm{q})\rangle
			\int d\bm{r}\mathcal{K}(\bm{r})e^{-i\bm{q}\cdot\bm{r}}=K_0(\hbar v_f)^2e^{-\frac{\xi^2\bm{q}^2}{2}}.
		\end{aligned}
	\end{equation}
	Based on the Born approximation, the self-energy  is 
	\begin{equation}\label{self-Born}
		\Sigma^{B}(\bm{k}s,E)=\sum_{s'}\int\frac{d^2\bm{k}'}{(2\pi)^2}\mathcal{K}(\bm{k}-\bm{k}')G^R_0(\bm{k}'s',E)\frac{1+ss'\cos\theta}{2}
	\end{equation}
	where $G^R_0(\bm{k}'s',E)=(E-s'\hbar v_fk'+i0^+)^{-1}$ is the Green's function of the clean system in the eigenstate basis, $\theta=\theta_{\bm{k}}-\theta_{\bm{k}'}$ is the angle between the wave vectors $\bm{k}$ and $\bm{k}'$. In the following, we separate the calculation of the self-energy function into real and imaginary parts, by using the Sokhotsky's formula,  $\frac{1}{x+i0^+}=\mathcal{P}\frac{1}{x}-i\pi\delta(x)$, where $\mathcal{P}$ denotes the Cauchy principal value.
	
	The imaginary part of the self-energy can be solved as
	\begin{equation}
		{\rm Im}\Sigma^{B}(\bm{k}s,E)=-\frac{K_0}{4}|E|e^{-\frac{\xi^2}{2}\left(k^2+\frac{E^2}{\hbar^2v^2_f}\right)}\left[I_0(\frac{\xi^2kE}{\hbar v_f})+sI_1(\frac{\xi^2kE}{\hbar v_f})\right]
	\end{equation}
	where $I_0(x)$ and $I_1(x)$ are the zero and first order modified Bessel functions of the first kind \cite{Gradshteyn,Abramowitz}.
	%We have expanded the results in power of energy $E$ and wave vector $k$ and omitted high order terms, since we focus on the region near the Dirac point (assuming $\frac{E}{\hbar v_f}\sim k\ll \frac{1}{d_0}$).
	Meanwhile, the real part of the self-energy is obtained as
	%\begin{equation}\begin{aligned}
	%	{\rm Re}\Sigma^{B}(\bm{k}s,E)=&\frac{K_0(\hbar v_f)^2}{8\pi^2}\sum_{s'}\mathcal{P}\int^{k_c}_0k'dk'\int^{\pi}_{-\pi}d\theta\frac{1}{E-s'\hbar v_fk'}\\
	%	&\times e^{-\frac{\xi^2}{2}(k^2+k'^2-2kk'\cos\theta)}(1+ss'\cos\theta)
	%\end{aligned}\end{equation}
	%After carrying out the angular integral, we get
	\begin{equation}\begin{aligned}
			{\rm Re}\Sigma^{B}(\bm{k}s,E)=&\frac{K_0(\hbar v_f)^2}{4\pi}\mathcal{P}\int^{k_c}_{-k_c}dk'k'e^{-\frac{\xi^2}{2}(k^2+k'^2)}\frac{I_0(\xi^2kk')+sI_1(\xi^2kk')}{E-\hbar v_fk'}.
	\end{aligned}\end{equation}
	where $k_c$ is the ultraviolet momentum cutoff. Focusing on the Dirac physics ($\frac{|E|}{\hbar v_f}, k\ll \frac{1}{\xi}$), we obtain the self-energy in powers of $k$ and $E$,
	\begin{eqnarray}
		\label{Im-2}
		{\rm Im}\Sigma^{B}(\bm{k}s,E)&=&-\frac{K_0}{4}|E| + \cdots \\
		\label{Re-2}
		{\rm Re}\Sigma^{B}(\bm{k}s,E)&=&\frac{K_0}{2\pi}E\ln|\frac{E}{\hbar v_fk_c}|+\alpha E-\alpha s\hbar v_fk + \cdots
	\end{eqnarray}
	where $\cdots$ stands for terms in the order of $O(kE^2,k^2E,E^3)$ in imaginary part and $O(kE^2\ln E,k^2E\ln E,E^3\ln E)$ in the real part. 
	From this result, a momentum dependent term appears in $Re \Sigma^B$, controlled by a dimensionless coefficient
	\begin{equation}\label{alpha}
		\alpha=K_0\frac{\xi^2k^2_c}{8\pi}.
	\end{equation}
	Note that, if the on-shell approximation  ($E=E_{\bm{k}s}=s\hbar v_fk$) is allowed, the momentum dependent correction in Eq.~(\ref{Re-2}) vanishes, recovering the previous results \cite{Sarma2011,Shon,Adam}.
	
	Here we stress that in the literature, the Born approximation is widely applied together with the on-shell approximation, so that the momentum dependence of the self-energy is ignored \cite{Trushin,Noro,Adam}. Our findings show that in the low energy regime the on-shell approximation is not well justified, i.e. the momentum and energy dependence in the self-energy are decoupled as shown in Eqs.~(\ref{Im-2}) and (\ref{Re-2}). Moreover, we find that the momentum dependent terms are less important in the imaginary part of self-energy than those in the real part, since in the latter they are at least two orders larger than the leading term in the imaginary part. Furthermore, our results can recover the predictions in the literature made using the on-shell approximation in the regime far away from the Dirac point \cite{Adam,McCann,Nersesyan}, where the imaginary part of the self-energy is a constant independent of energy and momentum, and the real part of self-energy disappears (details are shown in Supplemental Material Sec.~S2~A). Third, it should be noted that the Born approximation is essentially valid under weak disorder conditions ($K_0\lesssim1$) since it corresponds to the first order of the self-energy perturbation expansion. In this respect, one may wonder how the higher order expansions influence this calculation. We address this by considering a self-consistent Born approximation, and we have also confirmed a momentum dependent self-energy function, when the on-shell approximation is abandoned (see Supplemental Material Sec.~S2~B). 
	Finally, to further improve these perturbation calculations, we perform a renormalization group analysis on our model (Eq. \ref{Hamiltonian0}), and we confirm the momentum-dependent self-energy function cannot be neglected near the Dirac point (supplementary materials Sec. S3).
	Taken as a whole, different methods including the perturbation theory based on the (self-consistent) Born's approximation and renormalization group calculation, all reach the same conclusion, which strongly suggests our theoretical findings are robust and universal.  
	
	%%%%%%%%%%%%%%%%%%%%%%%%%%%%%%%%
	\begin{figure}
		\centering
		\includegraphics[width=0.8\linewidth]{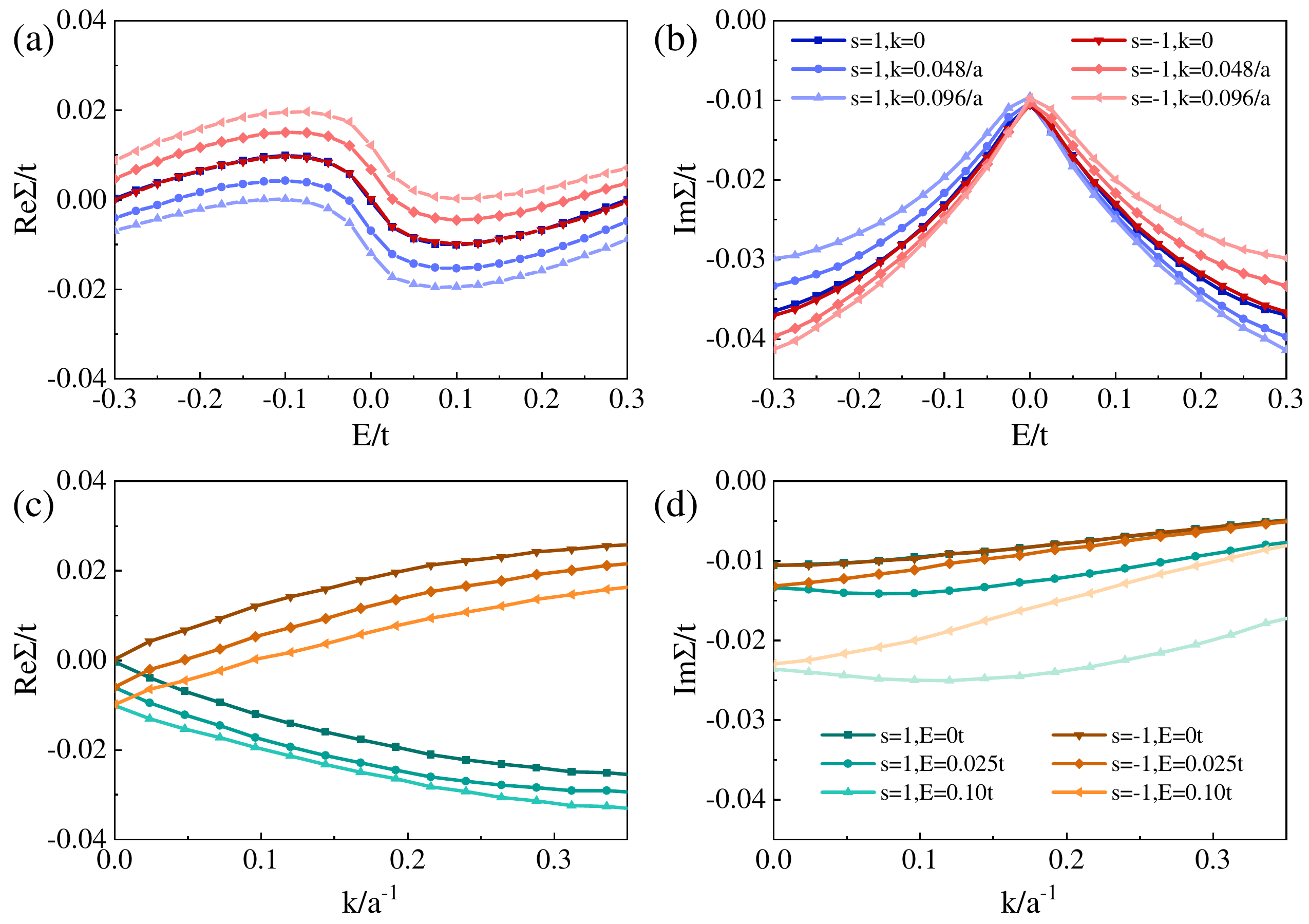}
		\caption{(a-b) Numerical results of $E$ vs ${\rm Re}\Sigma$ and $E$ vs ${\rm Im}\Sigma$ with different momenta $k$ and band index $s$. (c-d) Numerical results of $k$ vs ${\rm Re}\Sigma$ and $k$ vs ${\rm Im}\Sigma$ with different Fermi energies $E$ and band index $s$. Here we set an impurity concentration $n_{\text{imp}}=5\%$, correlation length $\xi=3.6a$, and impurity strength $u_0=0.16t$. Plots of $E$ vs $\Sigma$ ($k$ vs $\Sigma$) with more momentum values (energy values) are shown in the Supplemental Material Sec.~S3.}
		\label{fig:self-k-e}
	\end{figure}
	%%%%%%%%%%%%%%%%%%%%%%%%%%%%%%%%

	~\\
	\noindent\textbf{Numerical Simulations}\\
	%Next we turn to numerical simulations by Lanczos recursive method \cite{Chen2020}, using a large supercell with $N=2400\times2400\times2$ atoms and periodic boundary condition. 
	%The long-ranged Gaussian potential is chosen by $U(\bm{r})=\sum_{n=i}^{N_{\text{imp}}}\pm u_0e^{-|\bm{r}-\bm{R}_n|^2/\xi^2}$, where $\bm{R}_n$ and $N_{\text{imp}}$ are the location and total number of impurity and impurities of $\pm u_0$ are randomly distributed with equal probability. 
	Next we turn to numerical simulations using the Lanczos recursive method in momentum space ~\cite{Chen2020}. The Green's function $g(\bm{k}s,E)$ of the disordered system in the eigenbasis is directly calculated using the Lanczos recursive method, and the self-energy is obtained by the Dyson equation $\Sigma(\bm{k}s,E)=g_0^{-1}(\bm{k}s,E)-g^{-1}(\bm{k}s,E)$. In order to reach a high-energy resolution and reduce the finite-size errors, we consider a large supercell with $N=2400\times2400\times2$ atoms.  The data shown below are averaged over $50$ random configurations.
	In the discrete lattice model, the dimensionless parameter $K_0$ in the correlation function Eq.~(\ref{correlation-function}) has different expressions for sharp ($\xi\ll a$) and smooth ($\xi\gg a$) potentials \cite{Noro,Fan,Klos,Rycerz}
	\begin{equation}\label{K0}
		K_0=\frac{n_{\text{imp}}u^2_0}{(\hbar v_f)^2}\times\left\{\begin{array}{ll}
			\frac{A_c}{2};&\xi\ll a\\
			\frac{2\pi^2}{A_c}\xi^4;&\xi\gg a
		\end{array}\right.
	\end{equation}
	where $a=1.42\mathring{\mathrm{A}}$ the carbon-carbon distance of graphene, $A_c=\frac{3}{2}a^2$ is the area of the primitive cell, and $n_{\text{imp}}=N_{\text{imp}}/N$ is the impurity concentration.
	
	In Fig.~\ref{fig:self-k-e}, we display the self-energy functions versus Fermi energy and momentum. Obviously, the self-energy depends not only on the Fermi energy, but also on the momentum. As shown in the $E-{\rm Re}\Sigma$ plane of Fig.~\ref{fig:self-k-e}~(a), the momentum dependence of the real part of the self-energy function is significant in the low-energy region and gradually reduces with increasing energy, while the situation is opposite for the imaginary part in the $E-{\rm Im}\Sigma$ plane of Fig.~\ref{fig:self-k-e}~(b). From the curves in the $k-{\rm Re}\Sigma$ plane of Fig.~\ref{fig:self-k-e}~(c), we see that the real part of the self-energy function near the Dirac point is linearly related to the momentum, and the coefficient of the two eigenbands $s=\pm1$ are opposite. These behaviors mean that the Born estimation [Eq.~(\ref{Re-2})] without the on-shell approximation is reasonable around the Dirac point, and therefore we can assume the self-energy as Eq.~(\ref{re-k-E}). 
	%Furthermore, using the relation $\hbar v_f=\frac{3}{2}at$, where $t$ stands for the nearest-neighbor hopping energy of graphene lattice, we can numerically evaluate $\alpha=\frac{2}{3}\frac{\Delta R/t}{a\Delta k }$ by momentum shift $\Delta k$ and the corresponding change of real part of self-energy function $\Delta R$. 
	
\begin{figure}
		\centering
		\includegraphics[width=1.0\linewidth]{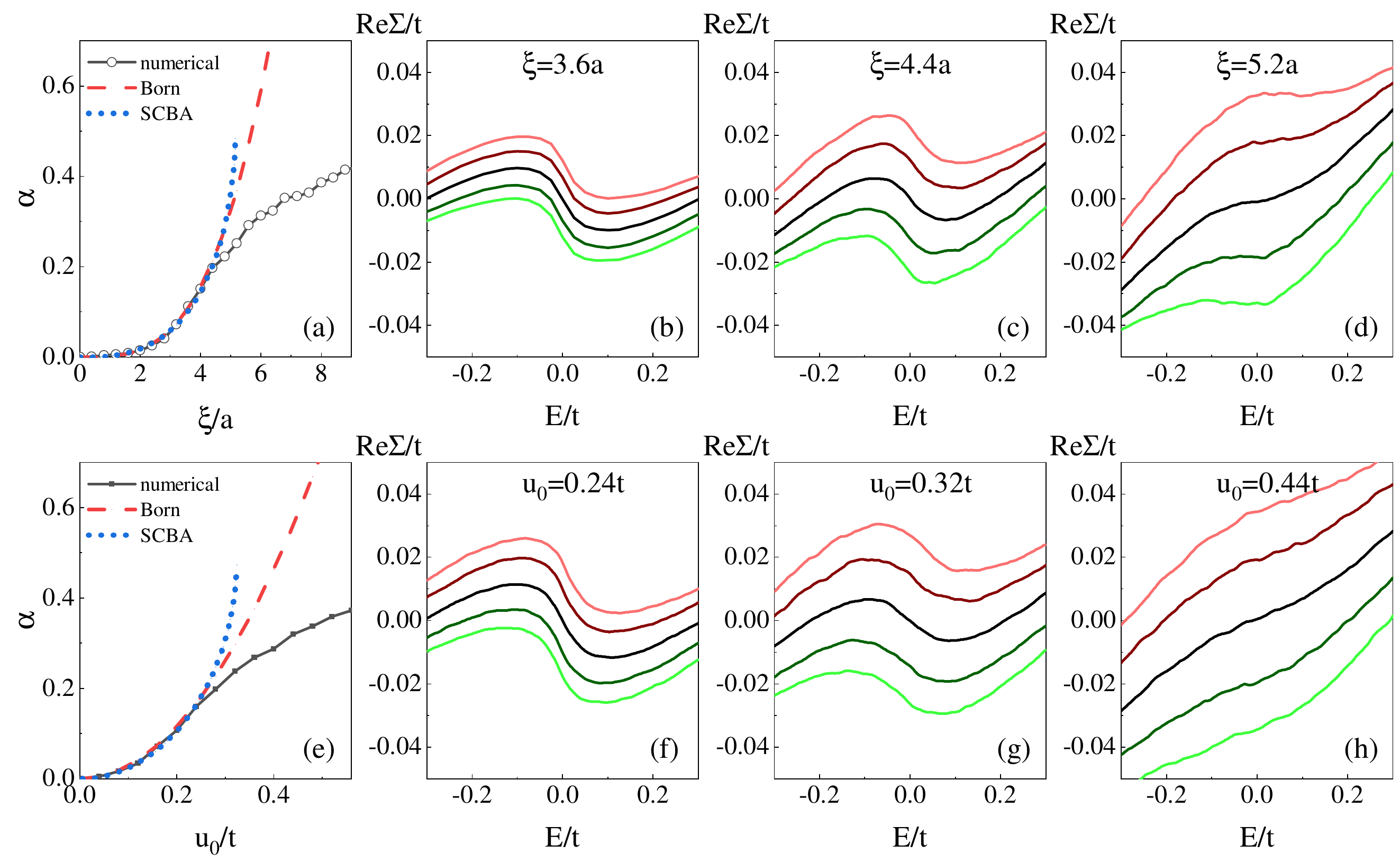}
		\caption{
			(a) $\alpha$ versus $\xi$ with $n_{\text{imp}}=5\%$ and $u_0=0.16t$. (b)-(d) Real part of the self-energy ${\rm Re}\Sigma$ for different $\xi=3.6a$ (b), $4.4a$ (c), and $5.2a$ (d). (e) $\alpha$ versus $u_0$ with $n_{\text{imp}}=5\%$ and $\xi=3.2a$. (f)-(h) Real part of the self-energy ${\rm Re}\Sigma$ for different $u_0=0.24t$ (f), $0.32t$ (g), and $0.44t$ (h). The dashed red lines (dotted blue lines) in (a,e) are fitting curves based on the Born approximation, which have $\alpha=1.6(\xi/a)^{3.3}\times10^{-3}$ and $\alpha=2.9(u_0/t)^2$. The dotted blue lines in (a,e) are fitting curves based on the self-consistent Born approximation, which have $\alpha=\frac{1}{2}-\sqrt{\frac{1}{4}-2.6(\xi/a)^{2.8}\times10^{-3}}$ and $\alpha=\frac{1}{2}-\sqrt{\frac{1}{4}-2.4(u_0/t)^2}$. In (b-d) and (f-h), the colors correspond to various momenta: Dirac point $k=0$ (black), $k=0.048/a$ and $s=-1$ (red), $k=0.096/a$ and $s=-1$ (light red), $k=0.048/a$ and $s=1$ (green), $k=0.096/a$ and $s=1$ (light green). 
		} 	\label{fig:alpha-re} 
	\end{figure}

	Figures~\ref{fig:alpha-re}~(a)-(d) exhibit the effect of the correlation length $\xi$ on the momentum dependence of the real part of the self-energy. 
	%The squares connected by solid line in Fig.~\ref{fig:alpha-re}~(a) are the numerical results of parameter $\alpha$ versus correlation length $\xi$. 
	When the disorder strength is small, $\alpha$ has a positive correlation with $\xi$, which can be fitted as $\alpha=1.6(\xi/a)^{3.3}\times10^{-3}$. This result is roughly consistent with the Born approximation which predicts: $\alpha\propto\xi^2$ in the limit $\xi\ll a$ and $\alpha\propto\xi^6$ in the limit $\xi\gg a$ according to Eqs.~(\ref{alpha}) and (\ref{K0}). Upon increasing $\xi$, the value of $\alpha$ tends to saturate and eventually decreases. This behavior indicates that the Born approximation tends to be invalid away from the weak-scattering limit, since the increase of $\xi$ also enhances the disorder strength $K_0$.
	
	Figures~\ref{fig:alpha-re}~(f)-(h) show the effect of the impurity strength $u_0$. Similar to the correlation length, we fit the relation of $\xi$ and $\alpha$ in the weak disorder regime, and obtain $\alpha=2.9(u_0/t)^{2}$. This result is in agreement with the prediction from the Born approximation: $\alpha\propto K_0\propto u_0^2$. 
	%Equivalent to the $\xi$, a further increase of $u_0$ will also cause the system to enter the strong scattering limit.
	
	In addition, from Fig.~\ref{fig:alpha-re}~(b)-(d) and (f)-(h), we find that the momentum dependence of the self-energy gradually decreases at high energy, where the on-shell approximation tends to be valid. The numerical simulation in the limit $k, \frac{E}{\hbar v_f}\gg \frac{1}{\xi}$ is further addressed in the Supplemental Material Sec.~S3.

		%%%%%%%%%%%%%%%%%%%%%%%%%%%%%%%%%
	%%%%%%%%%%%%%%%%%%%%%%%%%%%%%%%%%
	
	%%%%%%%%%%%%%%%%%
	%{\it Summary.---}
	~\\
	\noindent\textbf{Discussion and Summary}\\
	We have shown that in the presence of a long-ranged disorder potential, the self-energy function of the Dirac electrons has a peculiar dependence on the momentum, and this momentum dependence is of paramount importance for the transport properties of graphene, i.e. it produces a non-universal minimal conductivity. Furthermore, we elaborate the origin of this momentum dependent correction analytically and uncover its dependence on the disorder potential, which is also verified by unbiased numerical simulations. 
	%{\color{red}Moreover, both our analytical and numerical results show that the on-shell approximation tends to be valid at the limit $k, \frac{E}{\hbar v_f}\gg \frac{1}{\xi}$, which means that  for the high energy region, our main result Eq.~(\ref{re-k-E}), a decoupled momentum dependence of self-energy function, becomes invalid, and the self-energy returns to the predictions in previous theories \cite{Sarma2011,Shon,Adam}.}
	
	In closing, we would like to make several remarks. First, our findings offer an intuitive way for understanding the existing numerical simulations \cite{Nomura,Klos,Bardarson} where Coulomb impurities enhance the minimal conductivity.
	We believe that the physics behind these numerical calculations is now clear, from clarifying the self-energy function. 
	Second, another main cautionary message is that the on-shell approximation is not well justified around the Dirac point.
	Third, both analytical and numerical results show that the on-shell approximation tends to be valid in the limit $k, \frac{E}{\hbar v_f}\gg \frac{1}{\xi}$, which means that in the high energy region the self-energy function becomes momentum independent, in line with previous theories \cite{Sarma2011,Shon,Adam}.
	Finally, %since long-range potentials produce sample-dependent minimal conductivity, this suggests that long-ranged scatterers are the dominant microscopic source of remnant disorder in ultrahigh-mobility graphene.
	this work implies that a consistent theory for the quantum transport of graphene: Two main experimental observations, i.e. the linear dependence of the conductivity on carrier density \cite{Nomura} and the sample-dependent minimal conductivity, can be well understood by the presence of long-ranged impurity potentials. %Especially,   
	%as found by Nomura and MacDonald \cite{Nomura}, the form of disorder potential is relevant away from the Dirac point. In contrast, 
	%as shown in this paper, as the wave-length of electron near the Dirac point is longer than the spatial range of random potentials, the detailed form of impurity potential is irrelevant.  

	%\noindent\textbf{Method}\\

	%%
	{\it Acknowledgments.---} We thank Michael Smidman for critical proof reading of the final version of the current manuscript.
	This work was supported by ``Pioneer" and ''Leading Goose" R\&D Program of Zhejiang (2022SDXHDX0005), National Natural Science Foundation of China (No.12047544, No.11874337) and the foundation of Westlake University.

	{\it Author contributions.---}
	W.C., Q.W.S and W.Z. conceived the idea. W.C and Y.S carried out the analytical derivation and numerical calculation. W.C., B.F., and W.Z drafted the manuscript with discussions.

	%%%%%%%%%%%%%%%%%%%%%%%%%%
	%\bibliography{dynamic}
	
	%

	%%%%%%%%%%%%%%%%%%%%%%%%%%
	\clearpage
	
	\begin{appendix}
	\begin{widetext}
		
	\textbf{Supplemental Materials for ``On the sample-dependent minimal conductivity in weakly disordered graphene"}

	This supplemental material includes additional calculations to support the discussion in the main text. In Sec. S1, we show the analytical derivation of dc conductivity and analysis of the minimal conductivity. In Sec. S2, we compute the self-energy function of graphene in the presence of long-ranged Gaussian potential using the Born approximation and self-consistent Born approximation. We also treat the screened Coulomb potential and obtain the very similar expression for the self-energy function. In Sec. S3, we provide a way to understand our main conclusion from a different angle based on the renormalization group analysis. In Sec. S4, we display more numerical results of the self-energy function for both long-range Gaussian and Coulomb potentials. In Sec. S5, we exhibit the calculation of bubble diagram and vertex correction of dc conductivity in detail.
	
	\tableofcontents

	\section{Kubo formula for dc conductivity in terms of Green's function}
	The derivation of Eqs.~(5-7) in the main text, Kubo formula for the dc conductivity has been discussed in detail in some books of many-body quantum theory \cite{Mahan,Bruus}, so here we only briefly outline some significant steps. Based on the linear response theory, the dc conductivity given by
	\begin{equation}
		\sigma_{\alpha\beta}(E)=-\lim_{\omega\to0}\lim_{q\to0}\frac{{\rm Im}\mathcal{C}^R_{\alpha\beta}(\bm{q},\omega)}{\omega}
	\end{equation}
	where $\mathcal{C}^R_{\alpha\beta}(\bm{q},\omega)$ is a retarded current-current correlation function,
	\begin{equation}
		C^{R}_{\alpha\beta}(\bm{q},\omega)=-\frac{i}{\hbar}\frac{1}{\mathcal{V}}\int dt \theta(t-t')\langle[\hat{J}_{\alpha}(\bm{q},t),\hat{J}_{\beta}(-\bm{q},t')]\rangle e^{i\omega(t-t')}
	\end{equation}
	where $\hat{J}$ is the current operator, $\mathcal{V}$ is the area of sample and the bracket $\langle\cdots\rangle$ means the thermodynamic average over disorder. We use the Matsubara method to deal with the retarded correlation function, which indicates that the retarded correlation function is equivalent to the Matsubara function based on a analytical continuation: $\mathcal{C}(iq_n)\xrightarrow{iq_n=\omega+i\eta} C^R(\omega)$. 
	
	We start the calculation from the Matsubara current-current correlation function,
	\begin{equation}\begin{aligned}\label{eq:Mat-correlation-function}
			\mathcal{C}_{\alpha\beta}(\bm{q},iq_n)=&-\frac{1}{\hbar\mathcal{V}}\int_0^{\beta}d(\tau-\tau')\langle T_{\tau}\hat{J}_{\alpha}(\bm{q},\tau)\hat{J}_{\beta}(-\bm{q},\tau')\rangle e^{iq_n(\tau-\tau')}\\
			=&-\frac{1}{\hbar\mathcal{V}}\sum_{\bm{k}\bm{k}'}\int_0^{\beta}d(\tau-\tau')\langle T_{\tau}J_{\alpha,\bm{k},\bm{k}+\bm{q}}c_{\bm{k}}^{\dagger}(\tau)c_{\bm{k}+\bm{q}}(\tau)J_{\beta,\bm{k}',\bm{k}'-\bm{q}}c_{\bm{k}'}^{\dagger}(\tau')c_{\bm{k}'-\bm{q}}(\tau')\rangle e^{iq_n(\tau-\tau')}\\
			=&\frac{1}{\hbar\mathcal{V}}\sum_{\bm{k}\bm{k}'}\int_0^{\beta}d(\tau-\tau')e^{iq_n(\tau-\tau')}{\rm Tr}[ J_{\alpha,\bm{k},\bm{k}+\bm{q}}\mathcal{G}_{\bm{k}+\bm{q}}(\tau-\tau')J_{\beta,\bm{k}',\bm{k}'-\bm{q}}\mathcal{G}_{\bm{k}}(\tau'-\tau)\delta_{\bm{k}',\bm{k}+\bm{q}}]\\
			=&\frac{1}{\hbar\mathcal{V}}\sum_{\bm{k}}\int_0^{\beta}d(\tau-\tau')e^{iq_n(\tau-\tau')}{\rm Tr}[ J_{\alpha,\bm{k},\bm{k}+\bm{q}}\mathcal{G}_{\bm{k}+\bm{q}}(\tau-\tau')J_{\beta,\bm{k}+\bm{q},\bm{k}}\mathcal{G}_{\bm{k}}(\tau'-\tau)]\\
			=&\frac{1}{\hbar\mathcal{V}}\sum_{\bm{k}}\int_0^{\beta}d(\tau-\tau')\int_0^{\beta}d\tau_1\frac{1}{\beta}\sum_{ik_n}{\rm Tr}[ J_{\alpha,\bm{k},\bm{k}+\bm{q}}\mathcal{G}_{\bm{k}+\bm{q}}(\tau-\tau')e^{iq_n(\tau-\tau')}e^{ik_n(\tau-\tau')}J_{\beta,\bm{k}+\bm{q},\bm{k}}\mathcal{G}_{\bm{k}}(\tau_1)e^{ik_n\tau_1}]\\
			=&\frac{1}{\hbar\mathcal{V}\beta}\sum_{\bm{k},ik_n}{\rm Tr}[ J_{\alpha,\bm{k},\bm{k}+\bm{q}}\mathcal{G}_{\bm{k}+\bm{q},ik_n+iq_n}J_{\beta,\bm{k}+\bm{q},\bm{k}}\mathcal{G}_{\bm{k},ik_n}]\\
	\end{aligned}\end{equation}
	where $\beta=1/k_BT$, a closed fermion loop always gives a factor $-1$ and the trace of a product of Dirac matrices. $\mathcal{G}$ is Matsubara Green's function. The current operators are been expressed creation and annihilate operators: $\hat{J}_{\alpha}(\bm{q},\tau)=\sum_{\bm{k}}J_{\alpha,\bm{k},\bm{k}+\bm{q}}c_{\bm{k}}^{\dagger}(\tau)c_{\bm{k}+\bm{q}}(\tau)$. In addition, the identities $1=\int_0^{\beta}\delta(\tau_1-\tau'+\tau)d\tau_1$ and $\delta(\tau_1-\tau'+\tau)=\frac{1}{\beta}\sum_{ik_n}e^{ik_n(\tau_1-\tau'+\tau)}$ are applied. It is noticed that $ik_n=i(2n+1)\pi/\beta$ denotes Fermi frequency and $iq_n=i2n\pi/\beta$ denotes Boson frequency. 
	
	Next, we treat the summations of Matsubara Green’s functions with known branch cuts. According to the summation of $ik_n$, one can introduce a contour integral 
	\begin{equation}
		I=\oint_{\mathcal{C}_1+\mathcal{C}_2+\mathcal{C}_3}\frac{dz}{2\pi i}f(z)\mathcal{G}(z+iq_n)\mathcal{G}(z)
	\end{equation}
	where $f(z)=\frac{1}{e^{\beta z}+1}$ and the contours are shown in Fig.~\ref{fig:sketch-contour}, since for variable $z$ the branch cut (also singularity line) of $\mathcal{G}(z)$ is real axis, i.e. $z={\rm Re}z=\omega'$, and the branch cut of $\mathcal{G}(z+iq_n)$ is $z=\omega-iq_n$. On the one hand, we calculate $I$ by residue theorem
	\begin{equation}\label{eq-I-1}
		I=\sum_{ik_n}\lim_{z\to ik_n=i(2n+1)\pi/\beta}\frac{z-ik_n}{e^{ik_n\beta}+1}\mathcal{G}(ik_n+iq_n)\mathcal{G}(ik_n)=-\frac{1}{\beta}\sum_{ik_n}\mathcal{G}(ik_n+iq_n)\mathcal{G}(ik_n)
	\end{equation}
	On the other hand, we calculate $I$ by separating contour integrals
	\begin{equation}\begin{aligned}\label{eq-I-2}
			I=&\int_{-\infty}^{\infty}\frac{d\omega'}{2\pi i}f(z)\mathcal{G}(z+iq_n)\mathcal{G}(z)\bigg|_{z=\omega'-iq_n+i0^+}+\int_{\infty}^{-\infty}\frac{d\omega'}{2\pi i}f(z)\mathcal{G}(z+iq_n)\mathcal{G}(z)\bigg|_{z=\omega'-iq_n-i0^+}\\
			&+\int_{-\infty}^{\infty}\frac{d\omega'}{2\pi i}f(z)\mathcal{G}(z+iq_n)\mathcal{G}(z)\bigg|_{z=\omega'+i0^+}+\int_{\infty}^{-\infty}\frac{d\omega'}{2\pi i}f(z)\mathcal{G}(z+iq_n)\mathcal{G}(z)\bigg|_{z=\omega'-i0^+}\\
			=&\int_{-\infty}^{\infty}\frac{d\omega'}{2\pi i}f(\omega')[\mathcal{G}(\omega'+i0^+)\mathcal{G}(\omega'-iq_n)-\mathcal{G}(\omega'-i0^+)\mathcal{G}(\omega'-iq_n)+\mathcal{G}(\omega'+iq_n)\mathcal{G}(\omega'+i0^+)-\mathcal{G}(\omega'+iq_n)\mathcal{G}(\omega'-i0^+)]
	\end{aligned}\end{equation}
	Based on the comparison of the Eq.~(\ref{eq-I-1}) and Eq.~(\ref{eq-I-2}), we can rewrite Eq.~(\ref{eq:Mat-correlation-function}) as
	\begin{equation}\begin{aligned}
			\mathcal{C}_{\alpha\beta}(\bm{q},iq_n)=&\frac{-1}{\hbar\mathcal{V}}\sum_{\bm{k}}\int_{-\infty}^{\infty}\frac{d\omega'}{2\pi i}f(\omega'){\rm Tr}\left[J_{\alpha,\bm{k},\bm{k}+\bm{q}}\mathcal{G}_{\bm{k}+\bm{q},\omega'+i0^+}J_{\beta,\bm{k}+\bm{q},\bm{k}}\mathcal{G}_{\bm{k},\omega'-iq_n}-J_{\alpha,\bm{k},\bm{k}+\bm{q}}\mathcal{G}_{\bm{k}+\bm{q},\omega'-i0^+}J_{\beta,\bm{k}+\bm{q},\bm{k}}\mathcal{G}_{\bm{k},\omega'-iq_n}\right.\\
			&+\left.J_{\alpha,\bm{k},\bm{k}+\bm{q}}\mathcal{G}_{\bm{k}+\bm{q},\omega'+iq_n}J_{\beta,\bm{k}+\bm{q},\bm{k}}\mathcal{G}_{\bm{k},\omega'+i0^+}-J_{\alpha,\bm{k},\bm{k}+\bm{q}}\mathcal{G}_{\bm{k}+\bm{q},\omega'+iq_n}J_{\beta,\bm{k}+\bm{q},\bm{k}}\mathcal{G}_{\bm{k},\omega'-i0^+}\right]
	\end{aligned}\end{equation}
	
	Then, we do the analytical continuity $iq_n\to\omega+i0^+$ and get
	\begin{equation}\begin{aligned}
			C^R_{\alpha\beta}(\bm{q},\omega)=&\frac{-1}{\hbar\mathcal{V}}\sum_{\bm{k}}\int_{-\infty}^{\infty}\frac{d\omega'}{2\pi i}f(\omega'){\rm Tr}\left[J_{\alpha,\bm{k},\bm{k}+\bm{q}}\mathcal{G}_{\bm{k}+\bm{q},\omega'+i0^+}J_{\beta,\bm{k}+\bm{q},\bm{k}}\mathcal{G}_{\bm{k},\omega'-\omega-i0^+}-J_{\alpha,\bm{k},\bm{k}+\bm{q}}\mathcal{G}_{\bm{k}+\bm{q},\omega'-i0^+}J_{\beta,\bm{k}+\bm{q},\bm{k}}\mathcal{G}_{\bm{k},\omega'-\omega-i0^+}\right.\\
			&+\left.J_{\alpha,\bm{k},\bm{k}+\bm{q}}\mathcal{G}_{\bm{k}+\bm{q},\omega'+\omega+i0^+}J_{\beta,\bm{k}+\bm{q},\bm{k}}\mathcal{G}_{\bm{k},\omega'+i0^+}-J_{\alpha,\bm{k},\bm{k}+\bm{q}}\mathcal{G}_{\bm{k}+\bm{q},\omega'+\omega+i0^+}J_{\beta,\bm{k}+\bm{q},\bm{k}}\mathcal{G}_{\bm{k},\omega'-i0^+}\right]\\
			=&\frac{i}{\hbar\mathcal{V}}\sum_{\bm{k}}\int_{-\infty}^{\infty}\frac{d\omega'}{2\pi}f(\omega'){\rm Tr}\left[J_{\alpha,\bm{k},\bm{k}+\bm{q}}G^R_{\bm{k}+\bm{q}}(\omega')J_{\beta,\bm{k}+\bm{q},\bm{k}}G^A_{\bm{k}}(\omega'-\omega)-J_{\alpha,\bm{k},\bm{k}+\bm{q}}G^A_{\bm{k}+\bm{q}}(\omega')J_{\beta,\bm{k}+\bm{q},\bm{k}}G^A_{\bm{k}}(\omega'-\omega)\right.\\
			&+\left.J_{\alpha,\bm{k},\bm{k}+\bm{q}}G^R_{\bm{k}+\bm{q}}(\omega'+\omega)J_{\beta,\bm{k}+\bm{q},\bm{k}}G^R_{\bm{k}}(\omega')-J_{\alpha,\bm{k},\bm{k}+\bm{q}}G^R_{\bm{k}+\bm{q}}(\omega'+\omega)J_{\beta,\bm{k}+\bm{q},\bm{k}}G^A_{\bm{k}}(\omega')\right]\\
			=&\frac{i}{\hbar\mathcal{V}}\sum_{\bm{k}}\int_{-\infty}^{\infty}\frac{d\omega'}{2\pi}f(\omega'+\omega){\rm Tr}\left[J_{\alpha,\bm{k},\bm{k}+\bm{q}}G^R_{\bm{k}+\bm{q}}(\omega'+\omega)J_{\beta,\bm{k}+\bm{q},\bm{k}}G^A_{\bm{k}}(\omega')-J_{\alpha,\bm{k},\bm{k}+\bm{q}}G^A_{\bm{k}+\bm{q}}(\omega'+\omega)J_{\beta,\bm{k}+\bm{q},\bm{k}}G^A_{\bm{k}}(\omega')\right]\\
			&+f(\omega')\left[J_{\alpha,\bm{k},\bm{k}+\bm{q}}G^R_{\bm{k}+\bm{q}}(\omega'+\omega)J_{\beta,\bm{k}+\bm{q},\bm{k}}G^R_{\bm{k}}(\omega')-J_{\alpha,\bm{k},\bm{k}+\bm{q}}G^R_{\bm{k}+\bm{q}}(\omega'+\omega)J_{\beta,\bm{k}+\bm{q},\bm{k}}G^A_{\bm{k}}(\omega')\right]\\
			=&\frac{i}{\hbar\mathcal{V}}\sum_{\bm{k}}\int_{-\infty}^{\infty}\frac{d\omega'}{2\pi}\\
			&[f(\omega'+\omega)-f(\omega)]{\rm Tr}\left[J_{\alpha,\bm{k},\bm{k}+\bm{q}}G^R_{\bm{k}+\bm{q}}(\omega'+\omega)J_{\beta,\bm{k}+\bm{q},\bm{k}}G^A_{\bm{k}}(\omega')-J_{\alpha,\bm{k},\bm{k}+\bm{q}}G^A_{\bm{k}+\bm{q}}(\omega'+\omega)J_{\beta,\bm{k}+\bm{q},\bm{k}}G^A_{\bm{k}}(\omega')\right]\\
			&+f(\omega')\left[J_{\alpha,\bm{k},\bm{k}+\bm{q}}G^R_{\bm{k}+\bm{q}}(\omega'+\omega)J_{\beta,\bm{k}+\bm{q},\bm{k}}G^R_{\bm{k}}(\omega')-J_{\alpha,\bm{k},\bm{k}+\bm{q}}G^A_{\bm{k}+\bm{q}}(\omega'+\omega)J_{\beta,\bm{k}+\bm{q},\bm{k}}G^A_{\bm{k}}(\omega')\right]
	\end{aligned}\end{equation}
	
	\begin{figure}
		\centering
		\includegraphics[width=0.45\linewidth]{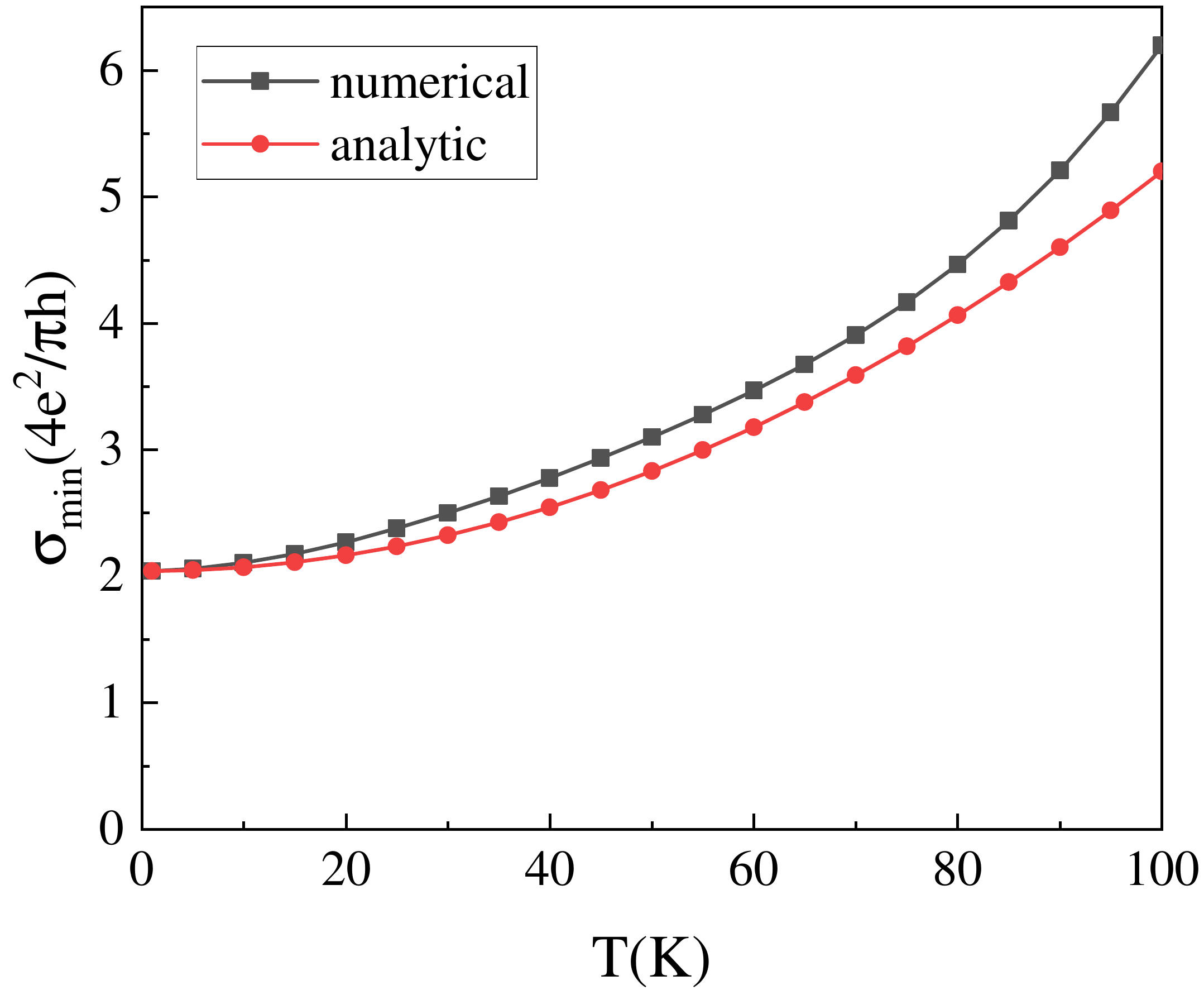}
		\caption{The temperature dependence of minimal conductivity obtained by numerical and analytical methods.}
		\label{fig:t-dependent}
	\end{figure}
	
	Finally, the dc longitudinal conductivity is obtained under the conditions: $\alpha=\beta=\hat{x}$, $\omega\to0$, $\bm{q}\to0$.
	\begin{equation}\begin{aligned}\label{eq:cond-finite-temperature}
			\sigma_{xx}(E)=&-\lim\limits_{\omega\to0}\frac{{\rm Im}C^R_{xx}(\bm{q}=0,\omega)}{\omega}\\
			=&\frac{e^2}{2\pi\hbar\mathcal{V}}\int_{-\infty}^{\infty}d\omega'\left[-\frac{\partial f(\omega')}{\partial\omega'}\right]{\rm ReTr}\left[v_{x}G^R(\omega')v_{x}G^A(\omega')-v_{x}G^R(\omega')v_{x}G^R(\omega')\right]
	\end{aligned}\end{equation}
	where the current is rewritten as $\bm{J}=e\bm{v}$. At the zero temperature, where $\frac{\partial f(\omega')}{\partial\omega'}=-\delta(\omega'-\frac{E}{\hbar})$ for $f(\omega)=\frac{1}{e^{(\hbar\omega-E)/k_BT}+1}$, the above expression can be further simplified to
	\begin{equation}
		\sigma_{xx}(E)=\frac{e^2\hbar}{2\pi\mathcal{V}}{\rm ReTr}\left[G^R(E)v_{x}G^A(E)v_{x}-G^R(E)v_{x}G^R(E)v_{x}\right]
	\end{equation}
	which corresponds to the Eqs.~(5-7) in the main text.

	While in the main text we only discuss the minimal conductivity at zero temperature, here we do some analysis of finite temperature effects. According to the Eq.~(\ref{eq:cond-finite-temperature}), the finite-temperature minimal conductivity can be obtained using the analytic Sommerfeld expansion or numerical integration. In the vicinity of the Dirac point, the temperature dependent dc conductivity can be evaluated as
	\begin{equation}\begin{aligned}\label{eq:minimal-temperature-evaluate}
			\sigma_{xx}(E\to0,T)\approx\frac{2e^2}{\pi h}\frac{1}{(1-\alpha)^2}\int_{-\infty}^{\infty}d\omega'\left[-\frac{\partial f(\omega')}{\partial\omega'}\right][2+(\frac{\omega'}{\eta_0})^2]=\frac{4e^2}{\pi h(1-\alpha)^2}+\frac{2e^2}{3h(1-\alpha)^2\eta_0^2}(k_BT)^2
	\end{aligned}\end{equation}
	where we have used a Sommerfeld expansion generating function $\int d\omega'\omega'^2\left[-\frac{\partial f(\omega')}{\partial\omega'}\right]=E^2+\frac{\pi^2}{3}(k_BT)^2$. This evaluation shows that the minimal conductivity at the finite temperature varies as $\Delta\sigma_{\text{min}}\propto T^2$. 
	
	Meanwhile, we show the numerical results of low-temperature dependence of minimal conductivity in Fig.~\ref{fig:t-dependent} and compare it with the above analytical results. In the numerical simulation, we consider the self-energy function obtained by Born approximation: $a=\omega'$, $\eta=\frac{K_0}{4}|\omega'|+\eta_0$ with the parameters $\eta_0=0.005$ and $K_0=0.3$.

	\begin{figure}
		\centering
		\includegraphics[width=0.6\linewidth]{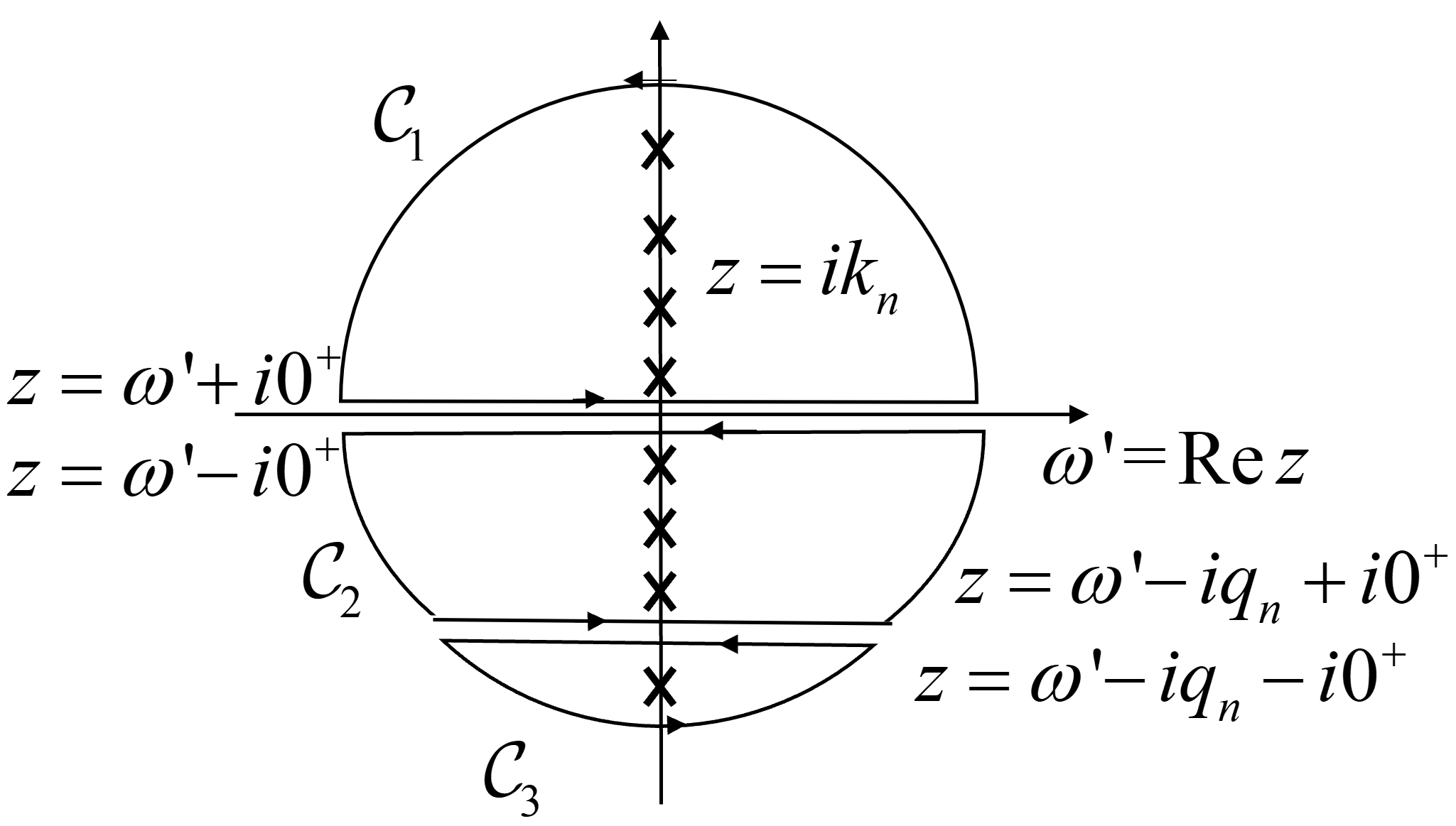}
		\caption{Contour integral with two branch cuts: $z={\rm Re}z=\omega$ and $z=\omega-i\Omega_n$.}
		\label{fig:sketch-contour}
	\end{figure}

	\section{Analytical Derivation of Self-Energy Function}
	\subsection{Long-ranged Gaussian Potential (Born approximation)}
	In this section, we evaluate the self-energy function in the presence of long-ranged Gaussian potential by the Born approximation:
		\begin{equation}\begin{aligned}
				\Sigma^{B}(\bm{k},E)=&\frac{1}{\mathcal{V}^2}\sum_{\bm{k}'}\overline{|V(\bm{k}-\bm{k}')|^2}U^{\dagger}_{\bm{k}}U_{\bm{k}'}G^R_0(\bm{k}',E)U^{\dagger}_{\bm{k}'}U_{\bm{k}}\\
				=&\frac{1}{2}\int\frac{d^2\bm{k}'}{(2\pi)^2}\mathcal{K}(\bm{k}-\bm{k}')
				\left(\begin{array}{cc}
					(g_0^++g_0^-)+\cos\theta(g_0^+-g_0^-)&i\sin\theta(g_0^+-g_0^-)\\
					-i\sin\theta(g_0^+-g_0^-)&(g_0^++g_0^-)-\cos\theta(g_0^+-g_0^-)
				\end{array}\right)
		\end{aligned}\end{equation}
	where $\overline{\cdots}$ represents the average of random configurations, $\theta=\theta_{\bm{k}}-\theta_{\bm{k}'}$ is the angle between momentum $\bm{k}$ and $\bm{k}'$, and $g_0^{\pm}=G^R_0(\bm{k}'\pm,E)=1/(E+i\eta-E_{\bm{k}'\pm})$ is the retarded Green's function for eigenstate with momentum $\bm{k}'$ and chiral $s=\pm1$. Based on the parity analysis, we find the integral of off-diagonal term is vanishing so that the first-order self-energy function in eigenstates basis can be expressed as
	\begin{equation}
		\Sigma^{B}(\bm{k}s,E)=\sum_{s'}\int\frac{d^2\bm{k}'}{(2\pi)^2}\mathcal{K}(\bm{k}-\bm{k}')G^R_0(\bm{k}'s',E)\frac{1+ss'\cos\theta}{2}
	\end{equation}
	which is the Eq.(14) in the main paper. After plugging the expression of $\mathcal{K}(\bm{k}-\bm{k}')$ into this equation, it is obtained
	\begin{equation}
		\Sigma^{B}(\bm{k}s,E)=K_0(\hbar v_f)^2\sum_{s'}\int\frac{d^2\bm{k}'}{(2\pi)^2}e^{-\frac{\xi^2(\bm{k}-\bm{k}')^2}{2}}G^R_0(\bm{k}'s',E)\frac{1+ss'\cos\theta}{2}
	\end{equation}
	For $E>0$, the imaginary part of the self-energy is
	\begin{equation}\begin{aligned}
			{\rm Im}\Sigma^{B}(\bm{k}s,E)=&K_0(\hbar v_f)^2\sum_{s'}\int\frac{d^2\bm{k}'}{(2\pi)^2}e^{-\frac{\xi^2(\bm{k}-\bm{k}')^2}{2}}\frac{1+ss'\cos\theta}{2}[-\pi\delta(E-s'\hbar v_fk')]\\
			=&-\frac{\pi K_0\hbar v_f}{2}\sum_{s'}\int\frac{k'dk'd\theta}{(2\pi)^2}e^{-\frac{\xi^2(k^2+k'^2-2kk'\cos\theta)}{2}}(1+ss'\cos\theta)\delta(\frac{E}{\hbar v_f}-s'k')\\
			=&-\frac{K_0}{8\pi}Ee^{-\frac{\xi^2}{2}\left(k^2+\frac{E^2}{\hbar^2v^2_f}\right)}\int^{\pi}_{-\pi} d\theta e^{\frac{\xi^2kE}{\hbar v_f}\cos\theta}(1+s\cos\theta)\\
			=&-\frac{K_0}{4}Ee^{-\frac{\xi^2}{2}\left(k^2+\frac{E^2}{\hbar^2v^2_f}\right)}\left[I_0(\frac{\xi^2kE}{\hbar v_f})+sI_1(\frac{\xi^2kE}{\hbar v_f})\right]
	\end{aligned}\end{equation}
	where $\tilde{E}=\frac{E}{\hbar v_f}$. In the final expression above, we have introduced modified Bessel functions of the first kind, whose integral representation is
	\begin{equation}
		I_{\alpha}(z)=\frac{1}{\pi}\int^{\pi}_0e^{z\cos\theta}\cos\alpha\theta d\theta-\frac{\sin\alpha\pi}{\pi}\int^{\infty}_0 e^{-z\cosh t-\alpha t}dt
	\end{equation}
	and has series expansion,
	\begin{equation}\label{I-small}
		I_{\alpha}(z)=\sum^{\infty}_{m=0}\frac{1}{m!\Gamma(m+\alpha+1)}\left(\frac{z}{2}\right)^{2m+\alpha},
	\end{equation}
	asymptotic expansion for large $z$,
	\begin{equation}\label{I-large}
		I_{\alpha}(z)=\frac{1}{\sqrt{2\pi z}\sqrt[4]{1+\frac{\alpha^2}{z^2}}}\exp\left[-\alpha{\rm arsinh}\left(\frac{\alpha}{z}\right)+z\sqrt{1+\frac{\alpha^2}{z^2}}\right]\left[1+o\left(\frac{1}{z\sqrt{1+\frac{\alpha^2}{z^2}}}\right)\right].
	\end{equation}

	Similarly, for $E<0$, the imaginary part of the self-energy is
	\begin{equation}\begin{aligned}
			{\rm Im}\Sigma^{B}(\bm{k}s,E)=&K_0(\hbar v_f)^2\sum_{s'}\int\frac{d^2\bm{k}'}{(2\pi)^2}e^{-\frac{\xi^2(\bm{k}-\bm{k}')^2}{2}}\frac{1+ss'\cos\theta}{2}[-\pi\delta(E-s'\hbar v_fk')]\\
			=&-\frac{\pi K_0\hbar v_f}{2}\sum_{s'}\int\frac{k'dk'd\theta}{(2\pi)^2}e^{-\frac{\xi^2(k^2+k'^2-2kk'\cos\theta)}{2}}(1+ss'\cos\theta)\delta(\frac{E}{\hbar v_f}-s'k')\\
			=&\frac{K_0}{8\pi}Ee^{-\frac{\xi^2}{2}\left(k^2+\frac{E^2}{\hbar^2v^2_f}\right)}\int^{\pi}_{-\pi} d\theta e^{-\frac{\xi^2kE\cos\theta}{\hbar v_f}}(1-s\cos\theta)\\
			=&\frac{K_0}{4}Ee^{-\frac{\xi^2}{2}\left(k^2+\frac{E^2}{\hbar^2v^2_f}\right)}\left[I_0(-\frac{\xi^2kE}{\hbar v_f})-sI_1(-\frac{\xi^2kE}{\hbar v_f})\right]
	\end{aligned}\end{equation}
	Combining two conditions, we can get
	\begin{equation}
		{\rm Im}\Sigma^{B}(\bm{k}s,E)=-\frac{K_0}{4}|E|e^{-\frac{\xi^2}{2}\left(k^2+\frac{E^2}{\hbar^2v^2_f}\right)}\left[I_0(\frac{\xi^2kE}{\hbar v_f})+sI_1(\frac{\xi^2kE}{\hbar v_f})\right]
	\end{equation}
	
	For the limit $k,\frac{E}{\hbar v_f}\ll\frac{1}{\xi}$, the above result can be expanded using Eq.(\ref{I-small}) as 
	\begin{equation}\begin{aligned}
			{\rm Im}\Sigma^{B}(\bm{k}s,E)=&-\frac{K_0}{4}|E|\left[1-\frac{\xi^2}{2}\left(k^2+\frac{E^2}{\hbar^2v^2_f}\right)+O(k^4,E^4,k^2E^2)\right]\left[1+s\frac{\xi^2kE}{2\hbar v_f}+O(k^2E^2)\right]\\
			=&-\frac{K_0|E|}{4}\left[1-\frac{\xi^2}{2}\left(k^2+\frac{E^2}{\hbar^2v^2_f}\right)+\frac{s\xi^2kE}{2\hbar v_f}\right]+O(k^4,E^4,k^2E^2)\\
			=&-\frac{K_0|E|}{4}+O(kE^2,k^2E,E^3)
	\end{aligned}\end{equation}
	which is the Eq.(17) in the main paper. 
	
	For the limit $k,\frac{E}{\hbar v_f}\gg\frac{1}{\xi}$, it has following asymptotic expression by using Eq.(\ref{I-large})
	\begin{equation}\begin{aligned}\label{im-high}
			{\rm Im}\Sigma^{B}(\bm{k}s,E)=&-\frac{K_0}{4}|E|e^{-\frac{\xi^2}{2}\left(k^2+\frac{E^2}{\hbar^2v^2_f}\right)}e^{\frac{\xi^2k|E|}{\hbar v_f}}\sqrt{\frac{\hbar v_f}{2\pi\xi^2k|E|}}\left[1+s{\rm sgn}(E)\right]\\
			=&-\frac{K_0}{4}|E|e^{-\frac{\xi^2}{2}\left(k-\frac{|E|}{\hbar v_f}\right)^2}\sqrt{\frac{\hbar v_f}{2\pi\xi^2k|E|}}\left[1+s{\rm sgn}(E)\right]\\
			\approx&-\frac{K_0(\hbar v_f)^2}{4\xi^2}\delta(E-E_{\bm{k}s})
	\end{aligned}\end{equation}
	where ${\rm sgn}(E)$ is sign function of $E$. The exponential decay term $e^{-\frac{\xi^2}{2}\left(k-\frac{|E|}{\hbar v_f}\right)^2}$ and the term $\left[1+s{\rm sgn}(E)\right]$ in the above at the second equation imply that the on-shell approximation ($E=E_{\bm{k}s}=s\hbar v_fk$) is valid in this condition based on the approximation $\lim\limits_{\sigma\to0}\frac{1}{\sqrt{2\pi\sigma}}\exp(-x^2/2\sigma)=\delta(x)$. The value of ${\rm Im}\Sigma^B$ we obtained under the on-shell approximation is a momentum and energy independent constant.
	
	The corresponding real part of self-energy function can be calculated by the Kramer-Kronig relation,
	\begin{equation}\begin{aligned}\label{k-k}
			{\rm Re}\Sigma^{B}(\bm{k}s,E)=&\frac{1}{\pi}\mathcal{P}\int^{E_c}_{-E_c}dE'\frac{{\rm Im}\Sigma^{B}(\bm{k}s,E')}{E'-E}\\
			=&\frac{1}{\pi}\mathcal{P}\int^{E_c}_{-E_c}dE'\frac{-\frac{K_0}{4}|E'|e^{-\frac{\xi^2}{2}\left(k^2+\frac{E'^2}{\hbar^2v^2_f}\right)}\left[I_0(\frac{\xi^2kE'}{\hbar v_f})+sI_1(\frac{\xi^2kE'}{\hbar v_f})\right]}{E'-E}
	\end{aligned}\end{equation}
	which is equivalent to the Eq.(16) in the main paper. 
	
	For the limit $k,\frac{E}{\hbar v_f}\ll\frac{1}{\xi}$, the corresponding real part of self-energy function is
	\begin{equation}\begin{aligned}
			{\rm Re}\Sigma^{B}(\bm{k}s,E)\approx&\frac{1}{\pi}\mathcal{P}\int^{E_c}_{-E_c}dE'\frac{-\frac{K_0}{4}|E'|[1-\frac{\xi^2}{2}\left(k^2+\frac{E'^2}{\hbar^2v^2_f}\right)]\left[1+\frac{1}{4}(\frac{\xi^2kE'}{\hbar v_f})^2+s\frac{\xi^2kE'}{2\hbar v_f}\right]}{E'-E}\\
			\approx&\frac{1}{\pi}\mathcal{P}\int^{E_c}_{-E_c}dE'\frac{-\frac{K_0}{4}|E'|\left[1+s\frac{\xi^2kE'}{2\hbar v_f}-\frac{\xi^2}{2}\left(k^2+\frac{E'^2}{\hbar^2v^2_f}\right)\right]}{E'-E}\\
			=&-\frac{K_0\hbar v_f}{4\pi}\mathcal{P}\left[\int^0_{-k_c}d\tilde{E}'\frac{-\tilde{E}'-\frac{s\xi^2k\tilde{E}'^2}{2}+\frac{\xi^2}{2}\tilde{E}'(k^2+\tilde{E}'^2)}{\tilde{E}'-\tilde{E}}+\int_0^{k_c}d\tilde{E}'\frac{\tilde{E}'+\frac{s\xi^2k\tilde{E}'^2}{2}-\frac{\xi^2}{2}\tilde{E}'(k^2+\tilde{E}'^2)}{\tilde{E}'-\tilde{E}}\right]\\
			=&-\frac{K_0\hbar v_f}{4\pi}\mathcal{P}\int_0^{k_c}d\tilde{E}'\left[\frac{-\tilde{E}'+\frac{s\xi^2k\tilde{E}'^2}{2}+\frac{\xi^2}{2}\tilde{E}'(k^2+\tilde{E}'^2)}{\tilde{E}'+\tilde{E}}+\frac{\tilde{E}'+\frac{s\xi^2k\tilde{E}'^2}{2}-\frac{\xi^2}{2}\tilde{E}'(k^2+\tilde{E}'^2)}{\tilde{E}'-\tilde{E}}\right]\\
			=&-\frac{K_0\hbar v_f}{4\pi}\mathcal{P}\int_0^{k_c}d\tilde{E}'\left(\frac{\tilde{E}'}{\tilde{E}'^2-\tilde{E}^2}2\tilde{E}+\frac{\tilde{E}'^3}{\tilde{E}'^2-\tilde{E}^2}s\xi^2k-\frac{\tilde{E}'(k^2+\tilde{E}'^2)}{\tilde{E}'^2-\tilde{E}^2}\xi^2\tilde{E}\right)\\
			=&\frac{K_0\hbar v_f}{2\pi}\left[\tilde{E}(\ln|\tilde{E}|-\ln|k_c|)-s\left(\frac{\xi^2k_c^2}{4}\right)k+\left(\frac{\xi^2k_c^2}{4}\right)\tilde{E}\right]+O(kE^2\ln E,k^2E\ln E,E^3\ln E)\\
			=&\frac{K_0}{2\pi}E\ln|\frac{E}{\hbar v_fk_c}|+\alpha E-s\alpha\hbar v_fk+O(kE^2\ln E,k^2E\ln E,E^3\ln E)
	\end{aligned}\end{equation}
	with
	\begin{equation}
		\alpha=K_0\frac{\xi^2k_c^2}{8\pi}
	\end{equation}
	so that we obtain the Eqs.(18) and (19) in the main paper. Here, $\tilde{E}=\frac{E}{\hbar v_f}$, $\tilde{E}'=\frac{E'}{\hbar v_f}$, and $k_c=\frac{E_c}{\hbar v_f}$. The modified Bessel functions are expanded as Eq.~(\ref{I-small}), and we keep the leading orders up to: $I_0(x)=1+\frac{x^2}{4}+o(x^3)$ and $I_1(x)=\frac{x}{2}+o(x^3)$.

	Then combining Eq.(\ref{im-high}) and (\ref{k-k}), we can obtain ${\rm Re}\Sigma^B$ at the limit $k,\frac{E}{\hbar v_f}\gg\frac{1}{\xi}$ as
	\begin{equation}\label{re-high}
		{\rm Re}\Sigma^B(\bm{k}s,E)\approx\frac{K_0(\hbar v_f)^2}{4\pi}\frac{E-E_{\bm{k}s}}{\xi^2(E-E_{\bm{k}s})^2+(\hbar v_f)^2}
	\end{equation}
	
	To sum up, we obtain the self-energy function for $\frac{|E|}{\hbar v_f}, k\ll \frac{1}{\xi}$
	\begin{eqnarray}
		\label{Im-2}
		{\rm Im}\Sigma^{B}(\bm{k}s,E)&=&-\frac{K_0}{4}|E| + \cdots \\
		\label{Re-2}
		{\rm Re}\Sigma^{B}(\bm{k}s,E)&=&\frac{K_0}{2\pi}E\ln|\frac{E}{\hbar v_fk_c}|+\alpha E-\alpha s\hbar v_fk + \cdots
	\end{eqnarray}
	and for $k,\frac{E}{\hbar v_f}\gg\frac{1}{\xi}$
	\begin{eqnarray}
		{\rm Im}\Sigma^{B}(\bm{k}s,E)&=&-\frac{K_0(\hbar v_f)^2}{4\xi^2}\delta(E-E_{\bm{k}s})\\
		{\rm Re}\Sigma^B(\bm{k}s,E)	&=& \frac{K_0(\hbar v_f)^2}{4\pi}\frac{E-E_{\bm{k}s}}{\xi^2(E-E_{\bm{k}s})^2+(\hbar v_f)^2}
	\end{eqnarray}

	\subsection{Long-ranged Gaussian Potential (SCBA)}
	Going beyond the Born approximation, we try to calculate the self-energy function using the self-consistent Born approximation (SCBA). It is expected that SCBA contains partial higher-order corrections in the perturbative expansion. Inspired by the Born approximation, we assume the self-energy function at the limit $k,\frac{E}{\hbar v_f}\ll\frac{1}{\xi}$ can be expressed in the form of
	\begin{equation}\label{eq:self-scba-assume}
		\Sigma(\bm{k}s,E)=\Sigma_1(E)-s\alpha\hbar v_fk+i\Sigma_2(E)=\tilde{\Sigma}(E)-s\alpha\hbar v_fk
	\end{equation}
	where $\tilde{\Sigma}(E)=\Sigma_1(E)+i\Sigma_2(E)$ is the part that depends only on energy. Thus, the Feynman diagram of SCBA is given by
	\begin{equation}\begin{aligned}\label{eq:self-scba-1}
			\Sigma(\bm{k}s,E)
			=&\sum_{s'}\int\frac{d^2\bm{k}'}{(2\pi)^2}\mathcal{K}(\bm{k}-\bm{k}')G(\bm{k}'s',E)\frac{1+ss'\cos\theta}{2}\\
			=&K_0(\hbar v_f)^2\sum_{s'}\int\frac{d^2\bm{k}'}{(2\pi)^2}e^{-\frac{\xi^2(\bm{k}-\bm{k}')^2}{2}}\frac{1}{E-E_{\bm{k}'s'}-\Sigma(\bm{k}'s',E)}\frac{1+ss'\cos\theta}{2}\\
			=&K_0(\hbar v_f)^2\sum_{s'}\int\frac{d^2\bm{k}'}{(2\pi)^2}e^{-\frac{\xi^2(\bm{k}-\bm{k}')^2}{2}}\frac{1}{E-s'(1-\alpha)\hbar v_fk'-\tilde{\Sigma}}\frac{1+ss'\cos\theta}{2}\\
			=&\frac{K_0(\hbar v_f)^2}{2}\int\frac{k'dk'd\theta}{(2\pi)^2}e^{-\frac{\xi^2(\bm{k}-\bm{k}')^2}{2}}\left[\frac{1+s\cos\theta}{(E-\tilde{\Sigma})-(1-\alpha)\hbar v_fk'}+\frac{1-s\cos\theta}{(E-\tilde{\Sigma})+(1-\alpha)\hbar v_fk'}\right]\\
			=&\frac{K_0(\hbar v_f)^2}{2\pi^2}\int_0^{k_c}k'dk'\int_0^{\pi}d\theta e^{-\frac{\xi^2(k^2+k'^2-2kk'\cos\theta)}{2}}\frac{(E-\tilde{\Sigma})+s\cos\theta(1-\alpha)\hbar v_fk'}{(E-\tilde{\Sigma})^2-(1-\alpha)^2(\hbar v_fk')^2}\\
			=&\frac{K_0(\hbar v_f)^2}{2\pi}\int_0^{k_c}k'dk' e^{-\frac{\xi^2(k^2+k'^2)}{2}}\left[\frac{(E-\tilde{\Sigma})}{(E-\tilde{\Sigma})^2-(1-\alpha)^2(\hbar v_fk')^2}I_0(kk'\xi^2)+\frac{s(1-\alpha)\hbar v_fk'}{(E-\tilde{\Sigma})^2-(1-\alpha)^2(\hbar v_fk')^2}I_1(kk'\xi^2)\right]\\
			=&\frac{K_0(\hbar v_f)^2}{2\pi}\int_0^{k_c}k'dk' [1+o(k^2\xi^2)]e^{-\frac{\xi^2k'^2}{2}}\frac{(E-\tilde{\Sigma})[1+o(k^2\xi^2)]+s(1-\alpha)\hbar v_fk'[\frac{kk'\xi^2}{2}+o(k^2\xi^2)]}{(E-\tilde{\Sigma})^2-(1-\alpha)^2(\hbar v_fk')^2}\\
			\approx&\frac{K_0(\hbar v_f)^2}{2\pi}\int_0^{k_c}k'dk'\frac{(E-\tilde{\Sigma})e^{-\frac{\xi^2k'^2}{2}}}{(E-\tilde{\Sigma})^2-(1-\alpha)^2(\hbar v_fk')^2}+s(1-\alpha)\hbar v_fk\frac{\frac{\xi^2k'^2}{2}e^{-\frac{\xi^2k'^2}{2}}}{(E-\tilde{\Sigma})^2-(1-\alpha)^2(\hbar v_fk')^2}\\
			=&\frac{K_0}{4\pi(1-\alpha)^2}\int_0^{\frac{\xi^2k_c^2}{2}}dx\frac{(E-\tilde{\Sigma})e^{-x}}{(E-\tilde{\Sigma})^2\tilde{\xi}^2-x}+s(1-\alpha)\hbar v_fk\frac{xe^{-x}}{(E-\tilde{\Sigma})^2\tilde{\xi}^2-x}\\
			=&\frac{K_0}{4\pi(1-\alpha)^2}\left[(E-\tilde{\Sigma})e^{-(E-\tilde{\Sigma})^2\tilde{\xi}^2}\left({\rm Ei}[(E-\tilde{\Sigma})^2\tilde{\xi}^2]-{\rm Ei}[(E-\tilde{\Sigma})^2\tilde{\xi}^2-\Lambda^2\tilde{\xi}^2]\right)-s(1-\alpha)\hbar v_fk(1-e^{-\Lambda^2\tilde{\xi}^2})\right.\\
			&\left.+s(1-\alpha)\hbar v_fk(E-\tilde{\Sigma})^2\tilde{\xi}^2e^{-(E-\tilde{\Sigma})^2\tilde{\xi}^2}\left({\rm Ei}[(E-\tilde{\Sigma})^2\tilde{\xi}^2]-{\rm Ei}[(E-\tilde{\Sigma})^2\tilde{\xi}^2-\Lambda^2\tilde{\xi}^2]\right)\right]\\
			\approx&\frac{K_0}{4\pi(1-\alpha)^2}\left\{(E-\tilde{\Sigma})\left[\ln\frac{(E-\tilde{\Sigma})^2}{(E-\tilde{\Sigma})^2-\Lambda^2}+\Lambda^2\tilde{\xi}^2\right]-s(1-\alpha)\hbar v_fk\left[\Lambda^2-(E-\tilde{\Sigma})^2\ln\frac{(E-\tilde{\Sigma})^2}{(E-\tilde{\Sigma})^2-\Lambda^2}\right]\tilde{\xi}^2\right\}\\
	\end{aligned}\end{equation}
	where $\tilde{\xi}^2=\frac{\xi^2}{2(\hbar v_f)^2(1-\alpha)^2}$, $\tilde{k}=s(1-\alpha)E_c$, $E_c=\hbar v_fk_c$, and $\Lambda=(1-\alpha)(\hbar v_fk_c)$. ${\rm Ei}(x)$ is exponential integral function and can be expanded as ${\rm Ei}(x)=\gamma+\ln x+x+o(x^2)$. The first term in square brackets is consistent with the short-range disorder when setting $\alpha=0$ and $\xi=0$, which can be solved in two limits $|E|\ll\Gamma_0$ and $|E|\gg\Gamma_0$, where $\Gamma_0=E_ce^{-2\pi/K_0}$ is an exponential small energy scale defined by the imaginary part of self-energy at Dirac point \cite{Ostrovsky}. Similarly, we can simplify the first term in square brackets in the last line of Eq.(\ref{eq:self-scba-1}) at two limits.
	
	At the limit $|E|\ll\Gamma_0$, we estimate
	\begin{equation}\begin{aligned}
			(E-\tilde{\Sigma})\ln\frac{(E-\tilde{\Sigma})^2}{(E-\tilde{\Sigma})^2-\Lambda^2}\approx&(E-\Sigma_1-i\Sigma_2)\ln\frac{\Sigma_2^2+2i\Sigma_2(E-\Sigma_1)}{\Lambda^2}
			=(E-\Sigma_1-i\Sigma_2)\left[\ln\frac{\Sigma_2^2}{\Lambda^2}+\ln(1+2i\frac{E-\Sigma_1}{\Sigma_2})\right]\\
			\approx&(E-\Sigma_1-i\Sigma_2)\left[\ln\frac{\Sigma_2^2}{\Lambda^2}+2i\frac{E-\Sigma_1}{\Sigma_2}\right]\\
			\approx&(E-\Sigma_1)\ln\frac{\Sigma_2^2}{\Lambda^2}+2(E-\Sigma_1)-i\Sigma_2\ln\frac{\Sigma_2^2}{\Lambda^2}
	\end{aligned}\end{equation} 
	and
	\begin{equation}\begin{aligned}
			-(E-\tilde{\Sigma})^2\ln\frac{(E-\tilde{\Sigma})^2}{(E-\tilde{\Sigma})^2-\Lambda^2}\approx&[2i(E-\Sigma_1)\Sigma_2+\Sigma_2^2]\left[\ln\frac{\Sigma_2^2}{\Lambda^2}+2i\frac{E-\Sigma_1}{\Sigma_2}\right]
			\approx\Sigma_2^2\ln\frac{\Sigma_2^2}{\Lambda^2}
	\end{aligned}\end{equation} 
	In this condition, the Eq.~(\ref{eq:self-scba-1}) can be further reduced as
	\begin{equation}\begin{aligned}\label{eq:self-scba-2}
			\Sigma(\bm{k}s,E)
			\approx\frac{K_0}{4\pi(1-\alpha)^2}\left\{(E-\Sigma_1)\ln\frac{\Sigma_2^2}{\Lambda^2}+2(E-\Sigma_1)-i\Sigma_2\ln\frac{\Sigma_2^2}{\Lambda^2}-s(1-\alpha)\hbar v_fk\Lambda^2\tilde{\xi}^2\right\}\\
	\end{aligned}\end{equation}
	The above self-consistent equation can be separated into three parts corresponding to the $\Sigma_1$, $i\Sigma_2$, and $-s\alpha\hbar v_fk$ parts in the self-energy function,
	\begin{equation}\begin{aligned}\label{eq:scba-separated-1}
			\Sigma_1=&\frac{K_0}{4\pi(1-\alpha)^2}\left[(E-\Sigma_1)\ln\frac{\Sigma_2^2}{\Lambda^2}+2(E-\Sigma_1)\right]\\
			%=&\frac{K_0}{4\pi(1-\alpha)^2}2(E-\Sigma_1)-(E-\Sigma_1)\\
			i\Sigma_2=&\frac{K_0}{4\pi(1-\alpha)^2}(-i\Sigma_2)\ln\frac{\Sigma_2^2}{\Lambda^2}\\
			-s\alpha\hbar v_fk=&\frac{K_0}{4\pi(1-\alpha)^2}[-s(1-\alpha)\hbar v_fk]\Lambda^2\tilde{\xi}^2
	\end{aligned}\end{equation}
	Solving the above self-consistent equations, we can approximately obtain
	\begin{equation}\begin{aligned}\label{eq:scba-solution-1}
			\Sigma_1=&-\frac{2\pi(1-\alpha)^2}{K_0}E\\
			\Sigma_2=&-\Lambda e^{-\frac{2\pi(1-\alpha)^2}{K_0}}\\
			\alpha=&\frac{1}{2}-\sqrt{\frac{1}{4}-K_0\frac{\xi^2k_c^2}{8\pi}}
	\end{aligned}\end{equation}
	
	At the limit $|E|\gg|{\rm Im}\Sigma|$, we estimate
	\begin{equation}
		(E-\tilde{\Sigma})\ln\frac{(E-\tilde{\Sigma})^2}{(E-\tilde{\Sigma})^2-\Lambda^2}\approx E\ln\frac{E^2e^{-i\pi{\rm sgn}(E)}}{\Lambda^2}=E\left[\ln\frac{E^2}{\Lambda^2}-i\pi{\rm sgn}(E)\right]=E\ln\frac{E^2}{\Lambda^2}-i\pi|E|
	\end{equation} 
	\begin{equation}\begin{aligned}
			-(E-\tilde{\Sigma})^2\ln\frac{(E-\tilde{\Sigma})^2}{(E-\tilde{\Sigma})^2-\Lambda^2}\approx&-[E^2-2iE\Sigma_2]\left[\ln\frac{E^2}{\Lambda^2}-i\pi{\rm sgn}(E)\right]
			\approx -E^2\ln\frac{E^2}{\Lambda^2}+i\pi E|E|
	\end{aligned}\end{equation} 
	In this condition, the Eq.~(\ref{eq:self-scba-1}) can be further reduced as
	\begin{equation}\begin{aligned}\label{eq:self-scba-3}
			\Sigma(\bm{k}s,E)
			\approx\frac{K_0}{4\pi(1-\alpha)^2}\left\{E\ln\frac{E^2}{\Lambda^2}-i\pi|E|+(E-i\Sigma_2)\Lambda^2\tilde{\xi}^2-s(1-\alpha)\hbar v_fk\Lambda^2\tilde{\xi}^2\right\}\\
	\end{aligned}\end{equation}
	Similarly, this self-consistent equation can be solved by separating into three parts
	\begin{equation}\begin{aligned}\label{eq:scba-separated-2}
			\Sigma_1=&\frac{K_0}{4\pi(1-\alpha)^2}(E\ln\frac{E^2}{\Lambda^2}+E\Lambda^2\tilde{\xi}^2)\\
			i\Sigma_2=&\frac{K_0}{4\pi(1-\alpha)^2}(-i\pi|E|-i\Sigma_2\Lambda^2\tilde{\xi}^2)\\
			s\alpha\hbar v_fk=&\frac{K_0}{4\pi(1-\alpha)^2}[-s(1-\alpha)\hbar v_fk]\Lambda^2\tilde{\xi}^2
	\end{aligned}\end{equation}
	Solving the above self-consistent equations, we can approximately obtain
	\begin{equation}\begin{aligned}\label{eq:scba-solution-2}
			\Sigma_1=&\frac{K_0}{4\pi(1-\alpha)^2}E\ln\frac{E^2}{\Lambda^2}+\frac{K_0\xi^2k_c^2}{8\pi(1-\alpha)^2}E\\
			\Sigma_2=&-\frac{\pi K_0}{4\pi(1-\alpha)^2}|E|\\
			\alpha=&\frac{1}{2}-\sqrt{\frac{1}{4}-K_0\frac{\xi^2k_c^2}{8\pi}}
	\end{aligned}\end{equation}
	Combining Eq.~(\ref{eq:scba-solution-1} and Eq.~(\ref{eq:scba-solution-2}, the resulting self-energy function obtained from SCBA is
	\begin{equation}
		\Sigma(\bm{k}s,E)=\left\{
		\begin{aligned}
			&-\frac{2\pi(1-\alpha)^2}{K_0}E-s\alpha\hbar v_fk-i\Lambda e^{-\frac{2\pi(1-\alpha)^2}{K_0}};&|E|\ll\Gamma_0\\
			&\frac{K_0}{4\pi(1-\alpha)^2}E\ln\frac{E^2}{\Lambda^2}+\frac{\alpha}{1-\alpha}E-s\alpha\hbar v_fk-i\frac{\pi K_0}{4\pi(1-\alpha)^2}|E|;&|E|\gg\Gamma_0
		\end{aligned}
		\right.
	\end{equation}
	with
	\begin{equation}
		\alpha=\frac{1}{2}-\sqrt{\frac{1}{4}-K_0\frac{\xi^2k_c^2}{8\pi}}.
	\end{equation}
	
	From the SCBA's results, we find that the dimensionless parameter $\alpha$, which controls the momentum linear dependent term in self-energy, has a upper limit $\frac{1}{2}$. %This is well consistent with the numerical results shown in Figs.~2(a) and 2(e) in the main text. 
	%Additionally, it recovers the result in the Born's approximation at $K_0\frac{\xi^2k_c^2}{8\pi}\ll\frac{1}{4}$. 
	Please note that this result naturally revert to the Born's result at weak disorder limit, $\alpha\xrightarrow{K_0\frac{\xi^2k_c^2}{8\pi}\ll\frac{1}{4}}K_0\frac{\xi^2k_c^2}{8\pi}$. To sum up, we prove that the self-energy function is momentum-dependent via the SCBA. 
	More Feynman diagrams beyond the SCBA is hard to calculate, which will be further validated by the numerical simulations.

	\subsection{Long-range Screened Coulomb Potential}\label{sec:Coulomb-self}
	For simplicity, most of our discussions in this paper are based on the random Gaussian potential. Here, we present a calculation using the long-ranged screened Coulomb potential. We will elucidate that, the main findings shown in this paper are robust, independent of the specific form of impurity potential.
	
	The long-range screened Coulomb potential has the form in real space as
	\begin{equation}
		V(\bm{r})=\pm\frac{e^2}{\kappa r}e^{-q_sr}
	\end{equation}
	where $\kappa$ is the static dielectric constant, scatters of $\pm$ are randomly distributed with equal probability, and $q_s$ is the Thomas-Fermi screening constant. It can be written in momentum space as
	\begin{equation}\begin{aligned}
			V(\bm{q})=&\int d\bm{r}V(\bm{r})e^{-i\bm{q}\cdot\bm{r}}
			=\pm\int d\bm{r}\frac{e^2}{\kappa r}e^{-q_sr}e^{-i\bm{q}\cdot\bm{r}}=\pm\int rdrd\theta\frac{e^2}{\kappa r}e^{-(q_s+iq\cos\theta)r}
			=\pm\frac{e^2}{\kappa}\int drd\theta e^{-(q_s+iq\cos\theta)r}\\
			=&\pm\frac{e^2}{\kappa}\int d\theta \frac{e^{-(q_s+iq\cos\theta)r}}{q_s+iq\cos\theta}\bigg|^{0}_{+\infty}
			=\pm\frac{e^2}{\kappa}\int d\theta\frac{1}{q_s+iq\cos\theta}=\pm\frac{e^2}{\kappa}\frac{2}{\sqrt{q^2_s+q^2}}\arctan(\sqrt{\frac{q_s-iq}{q_s+iq}}\tan\frac{\theta}{2})\bigg|^{\pi}_{-\pi}\\
			=&\pm\frac{e^2}{\kappa}\frac{2\pi}{\sqrt{q_s^2+q^2}}
	\end{aligned}\end{equation}
	
	Similar to the long-ranged Gaussian potential, we obtain the self-energy of long-range screened Coulomb potential by Born approximation
	\begin{equation}
		\Sigma(\bm{k}s,E)=\sum_{s'}\int\frac{d^2\bm{k}'}{(2\pi)^2}\frac{V^2(\bm{k}-\bm{k}')}{\mathcal{V}}G^R_0(\bm{k}'s',E)\frac{1+ss'\cos\theta}{2}
	\end{equation}
	After plugging the expression of $V(\bm{k}-\bm{k}')$ into this equation, it is obtained
	\begin{equation}\begin{aligned}
			\Sigma(\bm{k}s,E)=&\sum_{s'}\int\frac{d^2\bm{k}'}{(2\pi)^2}\frac{N_i}{\mathcal{V}}(\frac{2\pi e^2}{\kappa})^2\frac{1}{q_s^2+|\bm{k}-\bm{k}'|^2}G^R_0(\bm{k}'s',E)\frac{1+ss'\cos\theta}{2}\\
			=&\sum_{s'}\int\frac{d^2\bm{k}'}{(2\pi)^2}\frac{n_iV_0^2}{q_s^2+k^2+k'^2-2kk'\cos\theta}G^R_0(\bm{k}'s',E)\frac{1+ss'\cos\theta}{2}\\
	\end{aligned}\end{equation}
	where $N_i$ and $n_i=N_i/\mathcal{V}$ are the number and concentration of impurities and $V_0=\frac{2\pi e^2}{\kappa}$. Then we the above equation under the domain $\frac{E}{\hbar v_f}, k\ll q_s$.
	
	For $E>0$, the imaginary part of the self-energy is
	\begin{equation}\begin{aligned}
			{\rm Im}\Sigma(\bm{k}s,E)=&\sum_{s'}\int\frac{d^2\bm{k}'}{(2\pi)^2}\frac{n_iV_0^2}{q_s^2+k^2+k'^2-2kk'\cos\theta}\frac{1+ss'\cos\theta}{2}[-\pi\delta(E-s'\hbar v_fk')]\\
			=&-\frac{\pi n_iV_0^2}{2}\int\frac{k'dk'd\theta}{(2\pi)^2}\frac{1+s\cos\theta}{q_s^2+k^2+k'^2-2kk'\cos\theta}\delta(E-\hbar v_fk')\\
			=&-\frac{n_iV_0^2\tilde{E}}{8\pi\hbar v_f}\int d\theta\frac{1+s\cos\theta}{q_s^2+k^2+\tilde{E}^2-2k\tilde{E}\cos\theta}\\
			=&-\frac{n_iV_0^2\tilde{E}}{8\pi\hbar v_f}\left(-\frac{1}{2k\tilde{E}}\right)\int d\theta\frac{s(q_s^2+k^2+\tilde{E}^2-2k\tilde{E}\cos\theta)-[s(q_s^2+k^2+\tilde{E}^2)+2k\tilde{E}]}{q_s^2+k^2+\tilde{E}^2-2k\tilde{E}\cos\theta}\\
			=&\frac{n_iV^2_0}{16\pi\hbar v_f k}\left\{2\pi s-[s(q_s^2+k^2+\tilde{E}^2)+2k\tilde{E}]\int d\theta\frac{1}{q_s^2+k^2+\tilde{E}^2-2k\tilde{E}\cos\theta}\right\}\\
			=&\frac{n_iV^2_0}{8\hbar v_fk}\left[s-\frac{s(q_s^2+k^2+\tilde{E}^2)+2k\tilde{E}}{\sqrt{(q_s^2+k^2+\tilde{E}^2-2k\tilde{E})(q_s^2+k^2+\tilde{E}^2+2k\tilde{E})}}\right]\\
			=&\frac{n_iV^2_0}{8\hbar v_fk}s\left[1-\sqrt{\frac{q^2_s+(k+s\tilde{E})^2}{q^2_s+(k-s\tilde{E})^2}}\right]=\frac{n_iV^2_0}{8\hbar v_fk}s\left[1-\sqrt{1+\frac{4sk\tilde{E}}{q^2_s+(k-s\tilde{E})^2}}\right]\\
			\approx&-\frac{n_iV^2_0}{4\hbar v_f}\frac{\tilde{E}}{q^2_s+(k-s\tilde{E})^2}
			%=&-\frac{\pi n_iV_0^2}{2}\int\frac{k'dk'd\theta}{(2\pi)^2}\frac{1+\cos\theta}{q_s^2+k^2+k'^2-2kk'\cos\theta}\frac{\delta(k-k')}{\hbar v_f}\\
			%=&-\frac{n_iV_0^2k}{8\pi\hbar v_f}\int d\theta\frac{1+\cos\theta}{q_s^2+2k^2-2k^2\cos\theta}\\
			%=&-\frac{n_iV_0^2k}{8\pi\hbar v_f}[\frac{2\pi}{\sqrt{q_s^4+4q^2_sk^2}}+\frac{\pi}{k^2}(\frac{q_s^2+2k^2}{\sqrt{q_s^4+4q^2_sk^2}}-1)]\\
			%=&-\frac{n_iV_0^2}{8\hbar v_f}\frac{1}{k}(\sqrt{1+4(\frac{k}{q_s})^2}-1)
	\end{aligned}\end{equation}
	where $\tilde{E}=\frac{E}{\hbar v_f}$. 
	
	For $E<0$, the imaginary part of the self-energy is
	\begin{equation}\begin{aligned}
			{\rm Im}\Sigma(\bm{k}s,E)=&\sum_{s'}\int\frac{d^2\bm{k}'}{(2\pi)^2}\frac{n_iV_0^2}{q_s^2+k^2+k'^2-2kk'\cos\theta}\frac{1+ss'\cos\theta}{2}[-\pi\delta(E-s'\hbar v_fk')]\\
			=&-\frac{\pi n_iV_0^2}{2}\int\frac{k'dk'd\theta}{(2\pi)^2}\frac{1-s\cos\theta}{q_s^2+k^2+k'^2-2kk'\cos\theta}\delta(E+\hbar v_fk')\\
			=&\frac{n_iV_0^2\tilde{E}}{8\pi\hbar v_f}\int d\theta\frac{1-s\cos\theta}{q_s^2+k^2+\tilde{E}^2+2k\tilde{E}\cos\theta}\\
			\approx&\frac{n_iV^2_0}{4\hbar v_f}\frac{\tilde{E}}{q^2_s+(k-s\tilde{E})^2}
	\end{aligned}\end{equation}
	Combining two conditions, we can get
	\begin{eqnarray}
		{\rm Im}\Sigma(\bm{k}s,E)=-\frac{n_iV^2_0}{4\hbar v_f}\frac{|\tilde{E}|}{q^2_s+(k-s\tilde{E})^2}
	\end{eqnarray}
	
	Then the corresponding real part of self-energy function can be calculated by the Kramer-Kronig relation
	\begin{equation}\begin{aligned}
			{\rm Re}\Sigma(\bm{k}s,E)=&\frac{1}{\pi}\mathcal{P}\int^{E_c}_{-E_c}dE'\frac{{\rm Im}\Sigma(\bm{k}s,E')}{E'-E}=-\frac{n_iV^2_0}{4\pi\hbar v_f}\int^{\tilde{E}_c}_{-\tilde{E}_c}d\tilde{E}'\frac{1}{\tilde{E}'-\tilde{E}}\frac{|\tilde{E}'|}{q^2_s+(k-s\tilde{E}')^2}\\
			=&-\frac{n_iV^2_0}{4\pi\hbar v_f}\left\{\int^{\tilde{E}_c}_0d\tilde{E}'\frac{\tilde{E}'}{(\tilde{E}'-\tilde{E})[q^2_s+(k-s\tilde{E}')^2]}+\int_{-\tilde{E}_c}^0d\tilde{E}'\frac{-\tilde{E}'}{(\tilde{E}'-\tilde{E})[q^2_s+(k-s\tilde{E}')^2]}\right\}\\
			=&-\frac{n_iV^2_0}{4\pi\hbar v_f}\left\{\int^{\tilde{E}_c}_0d\tilde{E}'\frac{\tilde{E}'}{(\tilde{E}'-\tilde{E})[q^2_s+(k-s\tilde{E}')^2]}+\int^{\tilde{E}_c}_0d\tilde{E}'\frac{\tilde{E}'}{(-\tilde{E}'-\tilde{E})[q^2_s+(k+s\tilde{E}')^2]}\right\}\\
			=&-\frac{n_iV^2_0}{4\pi\hbar v_f}\int^{\tilde{E}_c}_0d\tilde{E}'\left\{\frac{\tilde{E}'}{(\tilde{E}'-\tilde{E})[q^2_s+(k-s\tilde{E}')^2]}-\frac{\tilde{E}'}{(\tilde{E}'+\tilde{E})[q^2_s+(k+s\tilde{E}')^2]}\right\}\\
			\approx&-\frac{n_iV^2_0}{4\pi\hbar v_fq^2_s}\int^{\tilde{E}_c}_0d\tilde{E}'\left\{\frac{\tilde{E}'}{\tilde{E}'-\tilde{E}}\left[1-\frac{(k-s\tilde{E}')^2}{q^2_s}\right]-\frac{\tilde{E}'}{\tilde{E}'+\tilde{E}}\left[1-\frac{(k+s\tilde{E}')^2}{q^2_s}\right]\right\}\\
			=&-\frac{n_iV^2_0}{4\pi\hbar v_fq^2_s}\int^{\tilde{E}_c}_0d\tilde{E}'\left[\left(\frac{\tilde{E}'}{\tilde{E}'-\tilde{E}}-\frac{\tilde{E}'}{\tilde{E}'+\tilde{E}}\right)(1-\frac{k^2}{q^2_s}-\frac{\tilde{E}'^2}{q^2_s})+\left(\frac{\tilde{E}'}{\tilde{E}'-\tilde{E}}+\frac{\tilde{E}'}{\tilde{E}'+\tilde{E}}\right)\frac{2sk\tilde{E}'}{q^2_s}\right]\\
			=&-\frac{n_iV^2_0}{2\pi\hbar v_fq^2_s}\int^{\tilde{E}_c}_0d\tilde{E}'\left[\frac{\tilde{E}'}{\tilde{E}'^2-\tilde{E}^2}\tilde{E}(1-\frac{k^2}{q^2_s})+\frac{\tilde{E}'^3}{\tilde{E}'^2-\tilde{E}^2}\frac{2sk}{q^2_s}-\frac{\tilde{E}'^3}{\tilde{E}'^2-\tilde{E}^2}\frac{\tilde{E}}{q^2_s}\right]\\
			=&-\frac{n_iV^2_0}{2\pi\hbar v_fq^2_s}\left\{\tilde{E}(\ln\tilde{E}_c-\ln\tilde{E})(1-\frac{k^2}{q^2_s})+\left[\frac{1}{2}\tilde{E}^2_c+\tilde{E}^2(\ln\tilde{E}_c-\ln\tilde{E})\right]\left(\frac{2sk}{q^2_s}-\frac{\tilde{E}}{q^2_s}\right)\right\}
	\end{aligned}\end{equation}
	where $E_c=\hbar v_fk_c$.
	
	Thus, if we consider the domain $\frac{E}{\hbar v_f}, k\ll q_s$, the self-energy function can be estimated as
	\begin{eqnarray}
		\label{screen-re}
		{\rm Re}\Sigma(\bm{k}s,E)&=&\frac{n_iV^2_0}{2\pi(\hbar v_f)^2q^2_s}E\ln|\frac{E}{E_c}|+\alpha E-s\alpha\hbar v_fk+O(kE^2\ln E,k^2E\ln E,E^3\ln E)\\
		\label{screen-im}
		{\rm Im}\Sigma(\bm{k}s,E)&=&-\frac{n_iV^2_0}{4(\hbar v_f)^2q^2_s}|E|+O(kE^2\ln E,k^2E\ln E,E^3\ln E)
	\end{eqnarray}
	where
	\begin{equation}
		\alpha=\frac{n_iV_0^2E^2_c}{2\pi(\hbar v_fq_s)^2}
	\end{equation}
	
	It can be seen from the above derivation that,  the self-energy function of long-range Screened Coulomb potential has same expression as that of long-ranged Gaussian potential expect for the value of $\alpha$. Physically, it can be understood from the following picture: Around the Dirac point, the form of random potential is not important, because the wave-length of electron is  longer than the spatial range of random potentials. So the different forms of random potentials give the similar results, as we shown in the paper. 
	However, away from the Dirac point, the form of disorder potential is relevant. 
	
	\newpage
	\section{Renormalization Group Analysis}
	
	In this section, we perform a Wilson's renormalization group calculation on the model that we studied in the main text. The purpose is two-fold. First, the renormalization group analysis could help to clarify some effect beyond SCBA  to the momentum dependent self-energy function. 
	Second, physically, the obtained flow equations clearly demonstrate the renormalized velocity together with the energy and disorder coupling constant, which provide a different angle to understand our main conclusion.  
	
	\subsubsection{Generating functional}
	
	At first, we expand the generating functional with cutoff prescription and disorder averaging. For simplicity, we set sources $\bar{\eta}=0$ and $\eta=0$.
	\begin{equation}\begin{aligned}\label{eq:RG-long-generating-start}
			\langle Z\rangle=&\left\langle\int D[\bar{\Psi},\Psi]_{\Lambda}\exp\left[i\int dt d^2\bm{x}\bar{\Psi}_{\bm{x},t}(i\partial_{t}+iv_f\nabla\cdot\bm{\sigma})\Psi_{\bm{x},\tau}\right]\exp\left[-i\int dt d^2\bm{x}V(\bm{x})\bar{\Psi}_{\bm{x},t}\Psi_{\bm{x},t}\right]\right\rangle\\
			=&\int D[\bar{\Psi},\Psi]_{\Lambda}\exp\left[\int d\tau d^2\bm{x}\bar{\Psi}_{\bm{x},\tau}(-\partial_{\tau}+iv_f\nabla\cdot\bm{\sigma})\Psi_{\bm{x},\tau}\right]\\
			&\left\langle1-\int d\tau d^2\bm{x}V(\bm{x})\bar{\Psi}_{\bm{x},\tau}\Psi_{\bm{x},\tau}+\frac{1}{2!}\int d\tau d^2\bm{x}d\tau'd^2\bm{x}'V(\bm{x})V(\bm{x}')\bar{\Psi}_{\bm{x},\tau}\Psi_{\bm{x},\tau}\bar{\Psi}_{\bm{x}',\tau'}\Psi_{\bm{x}',\tau'}\right.\\
			&-\frac{1}{3!}\int d\tau_1 d^2\bm{x}_1d\tau_2d^2\bm{x}_2d\tau_3d^2\bm{x}_3V(\bm{x}_1)V(\bm{x}_2)V(\bm{x}_3)\bar{\Psi}_{\bm{x}_1,\tau_1}\Psi_{\bm{x}_1,\tau_1}\bar{\Psi}_{\bm{x}_2,\tau_2}\Psi_{\bm{x}_2,\tau_2}\bar{\Psi}_{\bm{x}_3,\tau_3}\Psi_{\bm{x}_3,\tau_3}\\
			&+\left.\frac{1}{4!}\int d\tau_1 d^2\bm{x}_1d\tau_2d^2\bm{x}_2d\tau_3d^2\bm{x}_3d\tau_4d^2\bm{x}_4V(\bm{x}_1)V(\bm{x}_2)V(\bm{x}_3)V(\bm{x}_4)\bar{\Psi}_{\bm{x}_1,\tau_1}\Psi_{\bm{x}_1,\tau_1}\bar{\Psi}_{\bm{x}_2,\tau_2}\Psi_{\bm{x}_2,\tau_2}\bar{\Psi}_{\bm{x}_3,\tau_3}\Psi_{\bm{x}_3,\tau_3}\bar{\Psi}_{\bm{x}_4,\tau_4}\Psi_{\bm{x}_4,\tau_4}+\cdots\right\rangle\\
			=&\int D[\bar{\Psi},\Psi]_{\Lambda}\exp\left[\int\frac{d\omega}{2\pi}\frac{d^2\bm{k}}{(2\pi)^2}\bar{\Psi}_{\bm{k}\omega}(i\omega-v_f\bm{k}\cdot\bm{\sigma})\Psi_{\bm{k}\omega}\right]
			\left\{1+\frac{1}{2}\int d\tau d\tau'\frac{d^2\bm{q}d^2\bm{k}d^2\bm{k}'}{(2\pi)^6}\mathcal{K}(\bm{q})\bar{\Psi}_{\bm{k},\tau}\Psi_{\bm{k}-\bm{q},\tau}\bar{\Psi}_{\bm{k}',\tau'}\Psi_{\bm{k}'+\bm{q},\tau'}\right.\\
			&+\left.\frac{1}{8}\int d\tau_1 d\tau_2d\tau_3 d\tau_4 \frac{d^2\bm{q}d^2\bm{k}_1d^2\bm{k}_2}{(2\pi)^6}\frac{d^2\bm{q}'d^2\bm{k}_3d^2\bm{k}_4}{(2\pi)^6}\mathcal{K}(\bm{q})\mathcal{K}(\bm{q}')\bar{\Psi}_{\bm{k}_1,\tau_1}\Psi_{\bm{k}_1-\bm{q},\tau_1}\bar{\Psi}_{\bm{k}_2,\tau_2}\Psi_{\bm{k}_2+\bm{q},\tau_2}\bar{\Psi}_{\bm{k}_3,\tau_3}\Psi_{\bm{k}_3-\bm{q}',\tau_3}\bar{\Psi}_{\bm{k}_4,\tau_4}\Psi_{\bm{k}_4+\bm{q}',\tau_4}+\cdots\right\}\\
	\end{aligned}\end{equation}
	Notice that there are 3 combinations of $\langle V(\bm{x}_1)V(\bm{x}_2)\rangle\langle V(\bm{x}_3)V(\bm{x}_4)\rangle$. Here, we have transformed real time into imaginary time through Wick's rotation, $it\to\tau$, since it is convenient to use Matsubara Green functions in the following calculations. 
	
	\subsubsection{Momentum shell decomposition} 
	According to Wilson's approach, we divide the integration variables $\bar{\Psi}(\bm{k})$ and $\Psi(\bm{k})$ into two groups by a dimensionless variable $b>1$,
	\begin{equation}
		\bar{\Psi}(\bm{k})=\left\{
		\begin{aligned}
			&\bar{\Psi}^{<}(\bm{k});&0\leq|\bm{k}|<\Lambda/b;\\
			&\bar{\Psi}^{>}(\bm{k});&\Lambda/b\le|\bm{k}|<\Lambda;
		\end{aligned}
		\right.\ \ \ 
		\Psi(k)=\left\{
		\begin{aligned}
			&\Psi^{<}(\bm{k});&0\leq|\bm{k}|<\Lambda/b;\\
			&\Psi^{>}(\bm{k});&\Lambda/b\le|\bm{k}|<\Lambda;
		\end{aligned}
		\right.
	\end{equation}
	We replace the old $\bar{\Psi}$ and $\Psi$ with $\bar{\Psi}^<+\bar{\Psi}^>$ and $\Psi^<+\Psi^>$, and rewrite the generating functional as
	\begin{equation}\begin{aligned}\label{eq:RG-long-generating-decomposition}
			\langle Z\rangle
			=&\int D[\bar{\Psi}^<,\Psi^<]e^{S_0[\bar{\Psi}^<,\Psi^<]}\left\{\int D[\bar{\Psi}^>,\Psi^>]e^{S_0[\bar{\Psi}^>,\Psi^>]}\right\}
			\left\{1+\int d\tau d\tau'\frac{d^2\bm{q}d^2\bm{k}d^2\bm{k}'}{(2\pi)^6}\mathcal{K}(\bm{q})\bar{\Psi}^<_{\bm{k},\tau}\Psi^<_{\bm{k}-\bm{q},\tau}\Psi^<_{\bm{k}'+\bm{q},\tau'}\bar{\Psi}^<_{\bm{k}',\tau'}\right.\\
			&+\int d\tau d\tau'\frac{d^2\bm{q}d^2\bm{k}d^2\bm{k}'}{(2\pi)^6}\mathcal{K}(\bm{q})\bar{\Psi}^<_{\bm{k},\tau}\Psi^<_{\bm{k}'+\bm{q},\tau'}\langle\Psi^>_{\bm{k}-\bm{q},\tau}\bar{\Psi}^>_{\bm{k}',\tau'}\rangle\\
			&+\frac{1}{2}\int d\tau_1 d\tau_2d\tau_3 d\tau_4 \frac{d^2\bm{q}d^2\bm{k}_1d^2\bm{k}_2}{(2\pi)^6}\frac{d^2\bm{q}'d^2\bm{k}_3d^2\bm{k}_4}{(2\pi)^6}\mathcal{K}(\bm{q})\mathcal{K}(\bm{q}')\bar{\Psi}^<_{\bm{k}_1,\tau_1}\Psi^<_{\bm{k}_3-\bm{q}',\tau_3}\bar{\Psi}^<_{\bm{k}_4,\tau_4}\Psi^<_{\bm{k}_2+\bm{q},\tau_2}\langle\Psi^>_{\bm{k}_1-\bm{q},\tau_1}\bar{\Psi}^>_{\bm{k}_3,\tau_3}\rangle\langle\Psi^>_{\bm{k}_4+\bm{q}',\tau_4}\bar{\Psi}^>_{\bm{k}_2,\tau_2}\rangle\\
			&+\frac{1}{2}\int d\tau_1 d\tau_2d\tau_3 d\tau_4 \frac{d^2\bm{q}d^2\bm{k}_1d^2\bm{k}_2}{(2\pi)^6}\frac{d^2\bm{q}'d^2\bm{k}_3d^2\bm{k}_4}{(2\pi)^6}\mathcal{K}(\bm{q})\mathcal{K}(\bm{q}')\bar{\Psi}^<_{\bm{k}_1,\tau_1}\Psi^<_{\bm{k}_4+\bm{q}',\tau_4}\bar{\Psi}^<_{\bm{k}_2,\tau_2}\Psi^<_{\bm{k}_3-\bm{q}',\tau_3}\langle\Psi^>_{\bm{k}_1-\bm{q},\tau_1}\bar{\Psi}^>_{\bm{k}_4,\tau_4}\rangle\langle\Psi^>_{\bm{k}_2+\bm{q},\tau_2}\bar{\Psi}^>_{\bm{k}_3,\tau_3}\rangle\\
			&+\left.\int d\tau_1 d\tau_2d\tau_3 d\tau_4 \frac{d^2\bm{q}d^2\bm{k}_1d^2\bm{k}_2}{(2\pi)^6}\frac{d^2\bm{q}'d^2\bm{k}_3d^2\bm{k}_4}{(2\pi)^6}\mathcal{K}(\bm{q})\mathcal{K}(\bm{q}')\bar{\Psi}^<_{\bm{k}_1,\tau_1}\Psi^<_{\bm{k}_1-\bm{q},\tau_1}\bar{\Psi}^<_{\bm{k}_4,\tau_4}\Psi^<_{\bm{k}_3-\bm{q}',\tau_3}\langle\Psi^>_{\bm{k}_2+\bm{q},\tau_2}\bar{\Psi}^>_{\bm{k}_3,\tau_3}\rangle\langle\Psi^>_{\bm{k}_4+\bm{q}',\tau_4}\bar{\Psi}^>_{\bm{k}_2,\tau_2}\rangle+\cdots\right\}\\
			&=\int D[\bar{\Psi}^<,\Psi^<]e^{S_0[\bar{\Psi}^<,\Psi^<]}\left\{\int D[\bar{\Psi}^>,\Psi^>]e^{S_0[\bar{\Psi}^>,\Psi^>]}\right\}
			\left\{1+\int d\tau d\tau'\frac{d^2\bm{q}d^2\bm{k}d^2\bm{k}'}{(2\pi)^6}\mathcal{K}(\bm{q})\bar{\Psi}^<_{\bm{k},\tau}\Psi^<_{\bm{k}-\bm{q},\tau}\Psi^<_{\bm{k}'+\bm{q},\tau'}\bar{\Psi}^<_{\bm{k}',\tau'}\right.\\
			&-\int d\tau d\tau'\frac{d^2\bm{k}}{(2\pi)^2}\bar{\Psi}^<_{\bm{k},\tau}\Psi^<_{\bm{k},\tau'}\int\frac{d^2\bm{p}}{(2\pi)^2}\mathcal{K}(\bm{k}-\bm{p})\mathcal{G}^{>,0}(\bm{p},\tau-\tau')\\
			&+\frac{1}{2}\int d\tau_1 d\tau_2d\tau_3 d\tau_4 \frac{d^2\bm{q}d^2\bm{k}d^2\bm{k}'}{(2\pi)^6}\bar{\Psi}^<_{\bm{k},\tau_1}\Psi^<_{\bm{k}-\bm{q},\tau_3}\bar{\Psi}^<_{\bm{k}',\tau_4}\Psi^<_{\bm{k}'+\bm{q},\tau_2}\int\frac{d^2\bm{p}}{(2\pi)^2}\mathcal{K}(\bm{k}-\bm{p})\mathcal{K}(\bm{p}+\bm{q}-\bm{k})\mathcal{G}^{>,0}_{\bm{p},\tau_1-\tau_3}\mathcal{G}^{>,0}_{\bm{k}'+\bm{q}-\bm{k}+\bm{p},\tau_4-\tau_2}\\
			&+\frac{1}{2}\int d\tau_1 d\tau_2d\tau_3 d\tau_4 \frac{d^2\bm{q}d^2\bm{k}d^2\bm{k}'}{(2\pi)^6}\bar{\Psi}^<_{\bm{k},\tau_1}\Psi^<_{\bm{k}-\bm{q},\tau_4}\bar{\Psi}^<_{\bm{k}',\tau_2}\Psi^<_{\bm{k}'+\bm{q},\tau_3}\int\frac{d^2\bm{p}}{(2\pi)^2}\mathcal{K}(\bm{k}-\bm{p})\mathcal{K}(\bm{p}-\bm{k}+\bm{q})\mathcal{G}^{>,0}_{\bm{p},\tau_1-\tau_4}\mathcal{G}^{>,0}_{\bm{k}+\bm{k}'-\bm{p},\tau_2-\tau_3}\\
			&+\left.\int d\tau_1 d\tau_2d\tau_3 d\tau_4 \frac{d^2\bm{q}d^2\bm{k}d^2\bm{k}'}{(2\pi)^6}\bar{\Psi}^<_{\bm{k},\tau_1}\Psi^<_{\bm{k}-\bm{q},\tau_1}\bar{\Psi}^<_{\bm{k}',\tau_4}\Psi^<_{\bm{k}'+\bm{q},\tau_3}\int\frac{d^2\bm{p}}{(2\pi)^2}\mathcal{K}(\bm{q})\mathcal{K}(\bm{k}'-\bm{p})\mathcal{G}^{>,0}_{\bm{p},\tau_2-\tau_3}\mathcal{G}^{>,0}_{\bm{p}+\bm{q},\tau_4-\tau_2}+\cdots\right\}\\
	\end{aligned}\end{equation}
	where $S_0[\bar{\Psi}^<,\Psi^<]=\int\frac{d\omega}{2\pi}\frac{d^2\bm{k}}{(2\pi)^2}\bar{\Psi}^<_{\bm{k},\omega}(i\omega-v_f\bm{k}\cdot\bm{\sigma})\Psi^<_{\bm{k},\omega}$ is the unperturbed action in the momentum shell $|\bm{k}|<\Lambda/b$, $S_0[\bar{\Psi}^>,\Psi^>]=\int\frac{d\omega}{2\pi}\frac{d^2\bm{k}}{(2\pi)^2}\bar{\Psi}^>_{\bm{k},\omega}(i\omega-v_f\bm{k}\cdot\bm{\sigma})\Psi^>_{\bm{k},\omega}$ is the unperturbed action in the momentum shell $\Lambda/b<|\bm{k}|<\Lambda$, and $\langle\bar{\Psi}_1^>\Psi_2^>\rangle=\frac{\int D[\bar{\Psi}^>,\Psi^>]\Psi_1^>\bar{\Psi}_2^>e^{S_0[\bar{\Psi}^>,\Psi^>]}}{\int D[\bar{\Psi}^>,\Psi^>]e^{S_0[\bar{\Psi}^>,\Psi^>]}}=-\mathcal{G}^{>,0}_{12}$ denotes the unperturbed correction function for the fields $\bar{\Psi}^>$ and $\Psi^>$. The four terms containing integral of $\mathcal{G}^{>,0}$ in the brace correspond in turn to the four one-loop RG diagrams shown in Fig.~\ref{fig:rg-diagrams}. The diagram (a) is responsible for the renormalization of the energy  and velocity, while others are for disorder coupling. Additionally, diagram (a) has 2 degenerates due to the exchange $\tau\leftrightarrow\tau'$. Diagram (b) and (c) have 4 degenerates due to the exchanges $\tau_1\leftrightarrow\tau_2$ and $(\tau_1,\tau_2)\leftrightarrow(\tau_3,\tau_4)$. Diagram (d) has 8 degenerates due to the exchanges $\tau_1\leftrightarrow\tau_2$, $\tau_3\leftrightarrow\tau_4$ and $(\tau_1,\tau_2)\leftrightarrow(\tau_3,\tau_4)$. 
	
	\begin{figure}
		\centering
		\includegraphics[width=0.6\linewidth]{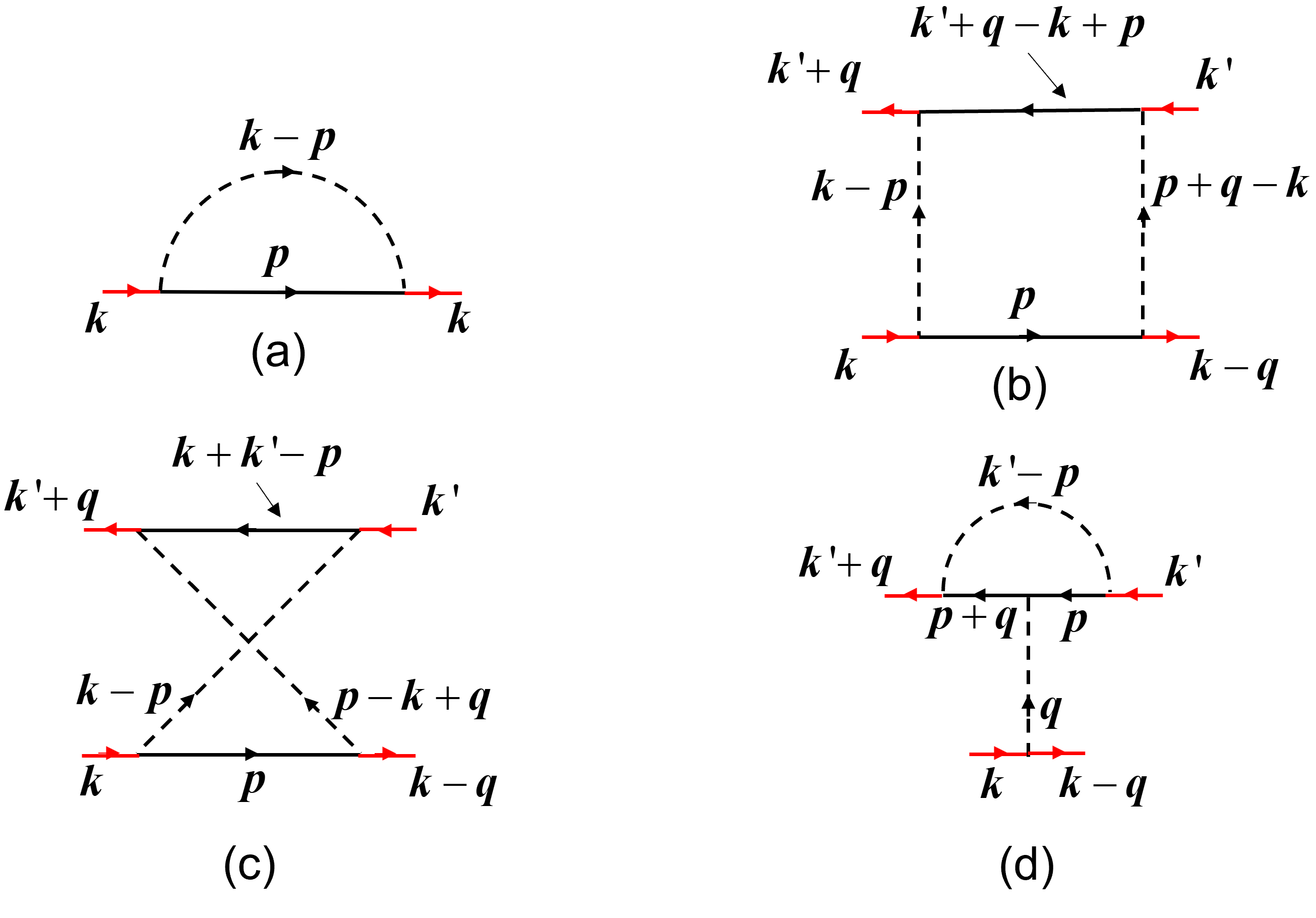}
		\caption{One loop RG diagrams responsible for the renormalization of (a) the energy and velocity, (b), (c), (d) the disorder coupling. Red lines denote external legs.}
		\label{fig:rg-diagrams}
	\end{figure}

	\subsubsection{Corrections of the energy, momentum, and disorder coupling}
	According to the Eq.~(\ref{eq:RG-long-generating-decomposition}), we can get the corrections of the energy, momentum, and disorder coupling after performing the integral over $\bar{\Psi}^>(\Psi^>)$. The corresponding four RG diagrams in Fig.~\ref{fig:rg-diagrams} are calculated one by one. When calculating the diagrams of the renormalization of disorder coupling, we assume that the momenta of the external lines are zero.
	\begin{equation}\begin{aligned}\label{eq:RG-a}
			I^{(a)}=&-\int_{[\Lambda/b,\Lambda]}\frac{d^2\bm{p}}{(2\pi)^2}\mathcal{K}(\bm{k}-\bm{p})\mathcal{G}^0(\bm{p},\omega)\\
			=&-K_0(\hbar v_f)^2\int_{[\Lambda/b,\Lambda]}\frac{d^2\bm{p}}{(2\pi)^2}e^{-\frac{\xi^2|\bm{k}-\bm{p}|^2}{2}}\frac{1}{i\omega-\hbar v_f\bm{p}\cdot\bm{\sigma}}
			=-K_0(\hbar v_f)^2\int_{[\Lambda/b,\Lambda]}\frac{d^2\bm{p}}{(2\pi)^2}e^{-\frac{\xi^2|\bm{k}-\bm{p}|^2}{2}}\frac{i\omega+\hbar v_f\bm{p}\cdot\bm{\sigma}}{\omega^2+(\hbar v_fp)^2}\\
			=&\frac{K_0(\hbar v_f)^2}{4\pi^2}\int_{-\pi}^{\pi}d\theta\int_{\Lambda/b}^{\Lambda}pdpe^{-\frac{\xi^2(k^2+p^2-2kp\cos\theta)}{2}}\left[\frac{i\omega}{\omega^2+(\hbar v_fp)^2}+\frac{\hbar v_fp\cos\theta}{\omega^2+(\hbar v_fp)^2}\frac{\bm{k}\cdot\bm{\sigma}}{k}\right]\\
			\approx&\frac{K_0}{4\pi^2}\int_{-\pi}^{\pi}d\theta\int_{\Lambda/b}^{\Lambda}dpe^{-\frac{\xi^2(k^2+p^2-2kp\cos\theta)}{2}}\left[\frac{i\omega}{p}+\cos\theta\frac{\hbar v_f\bm{k}\cdot\bm{\sigma}}{k}\right]\\
			=&\frac{K_0}{2\pi}\int_{\Lambda/b}^{\Lambda}dpe^{-\frac{\xi^2(k^2+p^2)}{2}}\left[I_0(kp\xi^2)\frac{i\omega}{p}+I_1(kp\xi^2)\frac{\hbar v_f\bm{k}\cdot\bm{\sigma}}{k}\right]
			\approx\frac{K_0}{2\pi}\int_{\Lambda/b}^{\Lambda}dp\left[\frac{i\omega}{p}+\frac{p\xi^2}{2}\hbar v_f\bm{k}\cdot\bm{\sigma}\right]\\
			=&\frac{K_0\ln b}{2\pi}i\omega+\frac{K_0\xi^2\Lambda^2}{8\pi}(1-b^{-2})\hbar v_f\bm{k}\cdot\bm{\sigma}\\
	\end{aligned}\end{equation}
	Here, we have assumed $k\xi,p\xi<1$, and the modified Bessel functions are expanded as $I_0(x)=1+\frac{x^2}{4}+o(x^3)$ and $I_1(x)=\frac{x}{2}+o(x^3)$. Meanwhile, $\theta$ is the angle between the momenta $\bm{k}$ and $\bm{p}$. The $\bm{p}\cdot\bm{\sigma}$ in the above derivation is transformed by
	\begin{equation}\begin{aligned}
			\bm{p}\cdot\bm{\sigma}=&p\cos\theta_{\bm{p}}\sigma_x+p\sin\theta_{\bm{p}}\sigma_y=p\cos(\theta_{\bm{k}}+\theta)\sigma_x+p\sin(\theta_{\bm{k}}+\theta)\sigma_y\\
			\to& \frac{p}{k}[k\cos\theta_{\bm{k}}\cos\theta\sigma_x+k\sin\theta_{\bm{k}}\cos\theta\sigma_y]\\
			=&\frac{p}{k}\cos\theta\bm{k}\cdot\bm{\sigma}
	\end{aligned}\end{equation}
	where the terms proportional to $\sin\theta$ are omitted based on the parity analysis of the integral.
	\begin{equation}\begin{aligned}\label{eq:RG-b}
			I^{(b)}=&\int_{[\Lambda/b,\Lambda]}\frac{d^2\bm{p}}{(2\pi)^2}\mathcal{K}(-\bm{p})\mathcal{K}(\bm{p})\mathcal{G}^0(\bm{p},\omega)\mathcal{G}^0(\bm{p},\omega)\\
			=&K_0^2(\hbar v_f)^4\int_{[\Lambda/b,\Lambda]}\frac{d^2\bm{p}}{(2\pi)^2}e^{-\xi^2p^2}\frac{1}{i\omega-\hbar v_f\bm{p}\cdot\bm{\sigma}}\frac{1}{i\omega-\hbar v_f\bm{p}\cdot\bm{\sigma}}
			=K_0^2(\hbar v_f)^4\int_{[\Lambda/b,\Lambda]}\frac{d^2\bm{p}}{(2\pi)^2}e^{-\xi^2p^2}\frac{(i\omega+\hbar v_f\bm{p}\cdot\bm{\sigma})^2}{[\omega^2+(\hbar v_fp)^2]^2}\\
			\approx&K_0^2(\hbar v_f)^2\int_{\Lambda/b}^{\Lambda}\frac{pdp}{2\pi}e^{-\xi^2p^2}\frac{1}{p^2}
			=\frac{K_0^2(\hbar v_f)^2}{4\pi}[{\rm Ei}(-\xi^2\Lambda^2)-{\rm Ei}(-\xi^2\Lambda^2/b^2)\\
			\approx&\frac{K_0^2(\hbar v_f)^2}{4\pi}[2\ln b+(1-b^{-2})\xi^2\Lambda^2]
	\end{aligned}\end{equation}
	\begin{equation}\begin{aligned}\label{eq:RG-c}
			I^{(c)}=&\int_{[\Lambda/b,\Lambda]}\frac{d^2\bm{p}}{(2\pi)^2}\mathcal{K}(-\bm{p})\mathcal{K}(\bm{p})\mathcal{G}^0(\bm{p},\omega)\mathcal{G}^0(-\bm{p},\omega)\\
			=&K_0^2(\hbar v_f)^4\int_{[\Lambda/b,\Lambda]}\frac{d^2\bm{p}}{(2\pi)^2}e^{-\xi^2p^2}\frac{1}{i\omega-\hbar v_f\bm{p}\cdot\bm{\sigma}}\frac{1}{i\omega+\hbar v_f\bm{p}\cdot\bm{\sigma}}
			=K_0^2(\hbar v_f)^4\int_{[\Lambda/b,\Lambda]}\frac{d^2\bm{p}}{(2\pi)^2}e^{-\xi^2p^2}\frac{(i\omega+\hbar v_f\bm{p}\cdot\bm{\sigma})(i\omega-\hbar v_f\bm{p}\cdot\bm{\sigma})}{[\omega^2+(\hbar v_fp)^2]^2}\\
			\approx&-K_0^2(\hbar v_f)^2\int_{\Lambda/b}^{\Lambda}\frac{pdp}{2\pi}e^{-\xi^2p^2}\frac{1}{p^2}
			=-\frac{K_0^2(\hbar v_f)^2}{4\pi}[{\rm Ei}(-\xi^2\Lambda^2)-{\rm Ei}(-\xi^2\Lambda^2/b^2)\\
			\approx&-\frac{K_0^2(\hbar v_f)^2}{4\pi}[2\ln b+(1-b^{-2})\xi^2\Lambda^2]
	\end{aligned}\end{equation}
	\begin{equation}\begin{aligned}\label{eq:RG-d}
			I^{(d)}=&\int_{[\Lambda/b,\Lambda]}\frac{d^2\bm{p}}{(2\pi)^2}\mathcal{K}(0)\mathcal{K}(-\bm{p})\mathcal{G}^0(\bm{p},\omega)\mathcal{G}^0(\bm{p},\omega)\\
			=&K_0^2(\hbar v_f)^4\int_{[\Lambda/b,\Lambda]}\frac{d^2\bm{p}}{(2\pi)^2}e^{-\xi^2p^2/2}\frac{1}{i\omega-\hbar v_f\bm{p}\cdot\bm{\sigma}}\frac{1}{i\omega-\hbar v_f\bm{p}\cdot\bm{\sigma}}
			=K_0^2(\hbar v_f)^4\int_{[\Lambda/b,\Lambda]}\frac{d^2\bm{p}}{(2\pi)^2}e^{-\xi^2p^2/2}\frac{(i\omega+\hbar v_f\bm{p}\cdot\bm{\sigma})^2}{[\omega^2+(\hbar v_fp)^2]^2}\\
			\approx&K_0^2(\hbar v_f)^2\int_{\Lambda/b}^{\Lambda}\frac{pdp}{2\pi}e^{-\xi^2p^2/2}\frac{1}{p^2}
			=\frac{K_0^2(\hbar v_f)^2}{4\pi}[{\rm Ei}(-\xi^2\Lambda^2/2)-{\rm Ei}(-\xi^2\Lambda^2/2b^2)\\
			\approx&\frac{K_0^2(\hbar v_f)^2}{4\pi}[2\ln b+(1-b^{-2})\xi^2\Lambda^2/2]
	\end{aligned}\end{equation}
	where ${\rm Ei}(x)$ is exponential integral function and can be expanded as ${\rm Ei}(-x)=\gamma+\ln x-x+o(x^2)$. The results of RG diagrams (b) [$I^{(b)}$] and (c) [$I^{(c)}$] cancel each other out.
	
	Plugging Eqs.~(\ref{eq:RG-a})-(\ref{eq:RG-d}) into Eq.~(\ref{eq:RG-long-generating-decomposition}), therefore, we can get the effective generating functional in the shell $|k|<\Lambda/b$ as
	\begin{equation}\begin{aligned}\label{eq:RG-generating-eff}
			\langle Z\rangle_{\text{eff}}=&\int D[\bar{\Psi}^<,\Psi^<]e^{S_0[\bar{\Psi}^<,\Psi^<]}
			\left\{1+\int d\tau d\tau'\frac{d^2\bm{q}d^2\bm{k}d^2\bm{k}'}{(2\pi)^6}\mathcal{K}(\bm{q})\bar{\Psi}^<_{\bm{k},\tau}\Psi^<_{\bm{k}-\bm{q},\tau}\Psi^<_{\bm{k}'+\bm{q},\tau'}\bar{\Psi}^<_{\bm{k}',\tau'}
			+\int\frac{d\omega}{2\pi}\frac{d^2\bm{k}}{(2\pi)^2}\bar{\Psi}^<_{\bm{k},\omega}\Psi^<_{\bm{k},\omega}(\Delta_Ei\omega+\alpha\hbar v_f\bm{k}\cdot\bm{\sigma})\right.\\
			&+\Delta_K\int d\tau_1 d\tau_2 \frac{d^2\bm{q}d^2\bm{k}d^2\bm{k}'}{(2\pi)^6}\bar{\Psi}^<_{\bm{k},\tau_1}\Psi^<_{\bm{k}-\bm{q},\tau_1}\bar{\Psi}^<_{\bm{k}',\tau_2}\Psi^<_{\bm{k}'+\bm{q},\tau_2}\\
			\approx&\int D[\bar{\Psi},\Psi]_{\Lambda/b}\\
			&\exp\left\{\int\frac{d\omega}{2\pi}\frac{d^2\bm{k}}{(2\pi)^2}\bar{\Psi}_{\bm{k},\omega}[(1+\Delta_E)i\omega-(1-\alpha)\hbar v_f\bm{k}\cdot\bm{\sigma}]\Psi_{\bm{k},\omega}+\int d\tau d\tau'\frac{d^2\bm{q}d^2\bm{k}d^2\bm{k}'}{(2\pi)^6}[\mathcal{K}(\bm{q})+\Delta_K]\bar{\Psi}_{\bm{k},\tau}\Psi_{\bm{k}-\bm{q},\tau}\Psi_{\bm{k}'+\bm{q},\tau'}\bar{\Psi}_{\bm{k}',\tau'}\right\}
	\end{aligned}\end{equation}
	with
	\begin{equation}
		\Delta_E=\frac{K_0\ln b}{2\pi};\ \ \ \alpha=\frac{K_0\xi^2\Lambda^2}{8\pi}(1-b^{-2});\ \ \ \ \Delta_K=\frac{K_0^2(\hbar v_f)^2}{4\pi}[2\ln b+(1-b^{-2})\xi^2\Lambda^2/2]
	\end{equation}
	The coefficient $\left\{\int D[\bar{\Psi}^>,\Psi^>]e^{S_0[\bar{\Psi}^>,\Psi^>]}\right\}$ is eliminated since it will be absorbed into the normalization of generating function.

	\subsubsection{Renormalization group flow} 
	Let us now rescale momenta and fields in the effective generating functional according to
	\begin{equation}
		\bm{k}'=b\bm{k};\ \ \ \bar{\Psi}'(\Psi')=b^{-3/2}\bar{\Psi}(\Psi)
	\end{equation}
	so that the momentum $\bm{k}'$ is integrated over $|k'|<\Lambda$. The rescaling of fields is to keep the free propagator unchanged. The rescaled effective generating functional is
	\begin{equation}\begin{aligned}
			\langle Z\rangle_{\text{eff}}&=\int D[\bar{\Psi}',\Psi']_{\Lambda}\\
			&\exp\left\{\int\frac{d\omega}{2\pi}\frac{d^2\bm{k}'}{(2\pi)^2}\bar{\Psi}'_{\bm{k}',\omega}[b(1+\Delta_E)i\omega-(1-\alpha)\hbar v_f\bm{k}'\cdot\bm{\sigma}]\Psi_{\bm{k}',\omega}+\int d\tau d\tau'\frac{d^2\bm{q}d^2\bm{k}d^2\bm{k}'}{(2\pi)^6}[\mathcal{K}(\bm{q})+\Delta_K]\bar{\Psi}^<_{\bm{k},\tau}\Psi^<_{\bm{k}-\bm{q},\tau}\Psi^<_{\bm{k}'+\bm{q},\tau'}\bar{\Psi}^<_{\bm{k}',\tau'}\right\}
	\end{aligned}\end{equation}
	which gives the transformation laws of energy, velocity and disorder
	\begin{equation}
		\left\{\begin{aligned}
			E'=&b(1+\Delta_E)E=b(1+\frac{K_0\ln b}{2\pi})E\\
			v_f'=&(1-\alpha)v_f=\left[1-\frac{K_0\xi^2\Lambda^2}{8\pi}(1-b^{-2})\right]v_f\\
			K_0'=&K_0+\frac{K_0^2}{8\pi}[4\ln b+(1-b^{-2})\xi^2\Lambda^2]
		\end{aligned}\right.
	\end{equation}
	
	where we only consider the leading term of disorder coupling $\mathcal{K}(\bm{q})\to K_0(\hbar v_f)^2$ and do the analytic continuation for energy, $i\omega\to E+i0^+$.
	
	Then we renormalization parameter $b=e^l\approx1+l$ and get the renormalization flows of energy, velocity and disorder
	\begin{equation}\label{eq:floweq}
		\left\{\begin{aligned}
			\frac{dE}{dl}=&(1+\frac{K_0}{2\pi})E\\
			\frac{dv_f}{dl}=&-\frac{K_0\xi^2\Lambda^2}{4\pi}v_f\\
			\frac{dK_0}{dl}=&\frac{K_0^2}{4\pi}(2+\xi^2\Lambda^2)
		\end{aligned}\right.
	\end{equation}

	Here we would like to provide some remarks. First,  by setting $\xi \rightarrow 0$, the flow equations Eq. \ref{eq:floweq} goes back to the existing results (e.g. Ref. \cite{Ostrovsky}). 
	That is, the renormalization of velocity is zero for short-ranged disorder potential. This is the main reason why the previous work overlooked the renormalization of velocity. 
	Second, as we shown here under the low energy condition ($k\xi, E\xi<1$), a long-ranged fluctuation $\xi \neq 0$ leads to renormalization of the velocity. Importantly, the effective velocity is reduced under the renormalization group flow.
	%{\color{blue} Physically it is reasonable, because the disorder scattering should hinder the electron transport. }
	Third, the renormalization group calculation is consistent with the calculation of self-energy function, because the momentum-dependent part of self-energy function, i.e. $\alpha \hbar v_f k$ in Eq. \ref{Re-2} or Eq. 18 in the main text, effectively reduces velocity in energy dispersion of electron.
	Taken all together, the independent renormalization group calculation further supports our conclusion shown in the main text.

	\newpage
	\section{Numerical Results of Self-Energy Function}
	
	\begin{figure}
		\centering
		\includegraphics[width=0.8\linewidth]{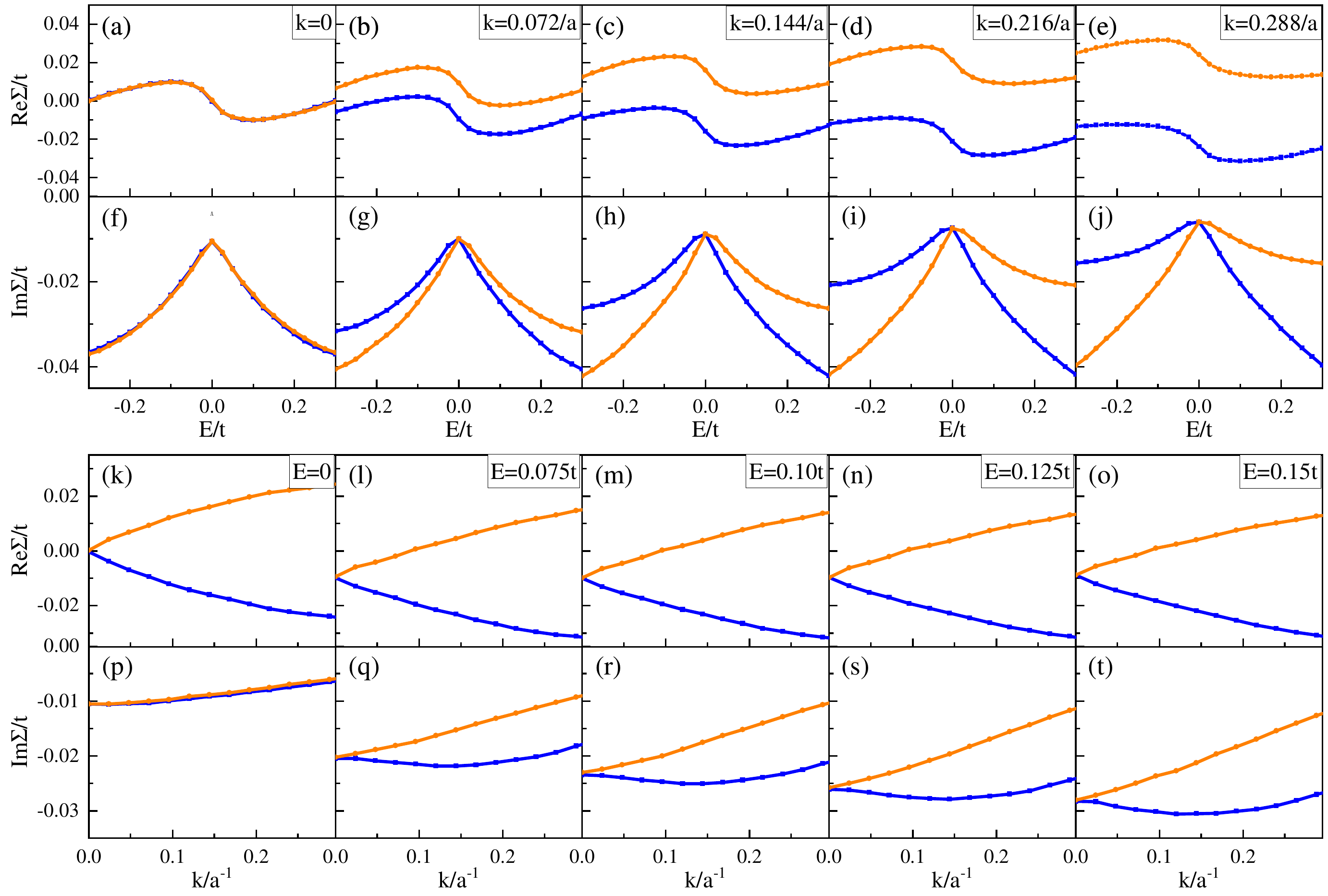}
		\caption{(a-e) Numerical results of ${\rm Re}\Sigma$ vs $E$ with different momentum. Insets in (a-e) are results but in a large energy range. (f-j) Numerical results of ${\rm Im}\Sigma$ vs $E$ with different momentum. (k-o) Numerical results of ${\rm Re}\Sigma$ vs $k$ with different Fermi energy. (p-t) Numerical results of ${\rm Im}\Sigma$ vs $k$ with different Fermi energy. Blue lines denotes subband with $s=1$ while orange lines denotes $s=-1$. Other parameters are the same as the FIG.~1 in the main paper: impurity concentration $n_{\text{imp}}=5\%$, correlation length $\xi=3.6a$, impurity strength $u_0=0.16t$.}
		\label{fig:self-k-e-2}
	\end{figure}
	\subsubsection{Long-ranged Gaussian Potential: $k,\frac{E}{\hbar v_f}\ll\frac{1}{\xi}$ regime}
	In the main paper, we have shown the numerical results of ${\rm Re}\Sigma/{\rm Im}\Sigma$ vs $E$ with momentum $k=0$, $0.048/a$ and $0.096/a$ and ${\rm Re}\Sigma/{\rm Im}\Sigma$ vs $k$ with Fermi energy $E=0t$, $0.025t$ and $0.050t$. Here, we show the evolution with more momentum and Fermi energy values in Fig.~\ref{fig:self-k-e-2}.

	\begin{figure}
		\centering
		\includegraphics[width=0.8\linewidth]{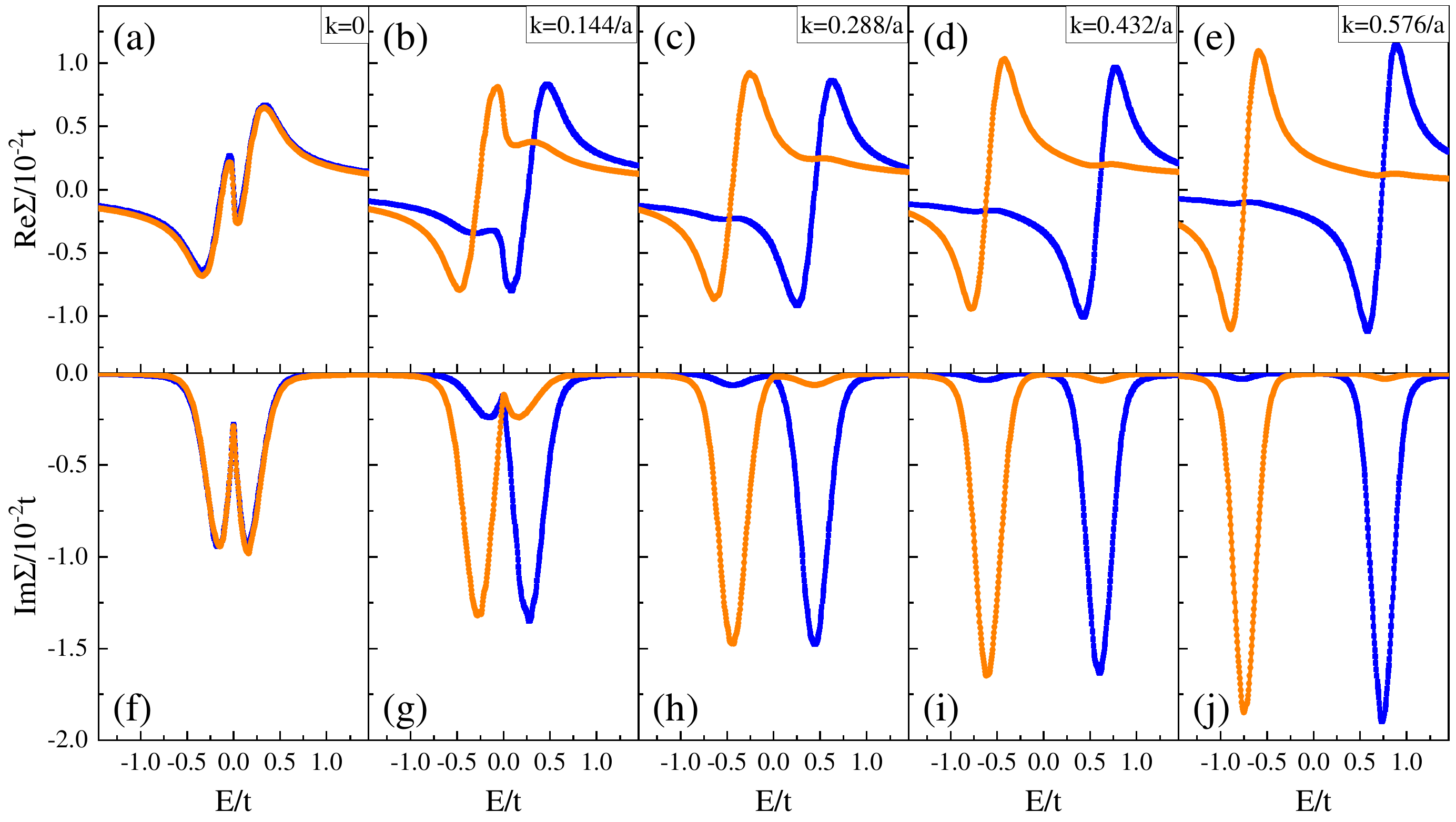}
		\caption{(Color online). (a-e) Numerical results of ${\rm Re}\Sigma$ vs $E$ with different momentum. (f-j) Numerical results of ${\rm Im}\Sigma$ vs $E$. Blue lines denotes subband with $s=1$ while orange lines denotes $s=-1$.  Other parameters: impurity concentration $n_{\text{imp}}=1\%$, correlation length $\xi=10a$, impurity strength $u_0=0.04t$.}
		\label{fig:self-k-e-large}
	\end{figure}
	\subsubsection{Long-ranged Gaussian Potential: $k,\frac{E}{\hbar v_f}\gg\frac{1}{\xi}$ regime}
	Since we have shown in the main paper the numerical self-energy function that satisfies $k,\frac{E}{\hbar v_f}\ll\frac{1}{\xi}$, here, we will show the corresponding result under the condition $k,\frac{E}{\hbar v_f}\gg\frac{1}{\xi}$. In the Fig. \ref{fig:self-k-e-large}, we display the self-energy functions versus Fermi energy with different momentum and versus momentum with different energy with impurity concentration $n_{\text{imp}}=1\%$, correlation length $\xi=10a$, impurity strength $u_0=0.04t$. For the very small Fermi energy and momentum in Figs.~\ref{fig:self-k-e-large}(a) and (f), where the condition $k,E\ll\frac{1}{\xi}$ still satisfied, the behavior of the real and imaginary parts of self-energy is qualitatively consistent with the results shown in Figs.~\ref{fig:self-k-e-2}(a) and (f). On the contrary, for the large Fermi energy and momentum, the imaginary part of self-energy tends to be a delta function which is same with the prediction from Born approximation Eq.~(\ref{im-high}). Similarly, the numerical result of real part of self-energy is also consistent with the analytical expression Eq.~(\ref{re-high}).
	
	\begin{figure}
		\centering
		\includegraphics[width=0.6\linewidth]{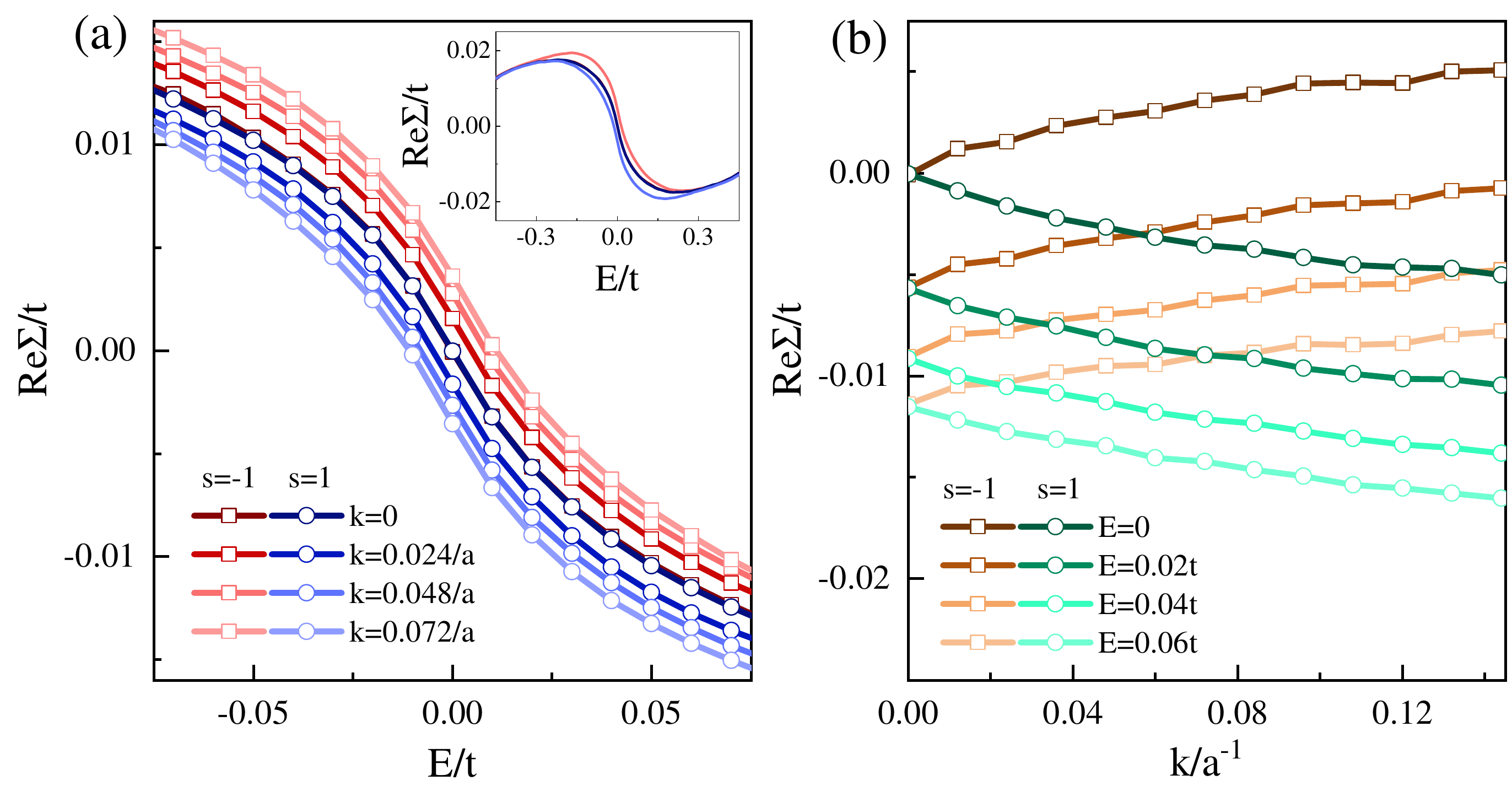}
		\caption{(Color online). (a) Numerical results of ${\rm Re}\Sigma$ vs $E$. (b) Numerical results of ${\rm Re}\Sigma$ vs $k$. Inset in (a) is same as (a) but in a large energy range and only keeping the data of $k=0$ and $k=0.072/a$. Other parameters: $n_i=0.05$, $V_0=0.1t$, and $q_s=0.1a$. In order to avoid the divergence at $r\to0$ in the discrete simulation, we set a minimum radius $r_c=0.01a$.}
		\label{fig:self-coulomb}
	\end{figure}
	\subsubsection{Long-range Screened Coulomb Potential: $k,\frac{E}{\hbar v_f}\ll q_s$ regime}
	
	In the Sec.~\ref{sec:Coulomb-self}, we have derived the self-energy function in the presence of long-range screened Coulomb potential based on the Born's approximation. We also calculate it by numerical simulation. In Fig.~\ref{fig:self-coulomb}, we show the numerical results of ${\rm Re}\Sigma$ vs $E$ and ${\rm Re}\Sigma$ vs $k$ near the Dirac point with parameters: $n_i=0.05$, $V_0=0.1t$, and $q_s=0.1a$. Additionally, in order to avoid the divergence at $r\to0$ in the discrete simulation, we set a minimum radius $r_c=0.01a$.

	\newpage
	\section{Kubo Formalism for Bubble Diagram and Vertex Correction of Conductivity}
	
	\begin{figure}
		\centering
		\includegraphics[width=0.7\linewidth]{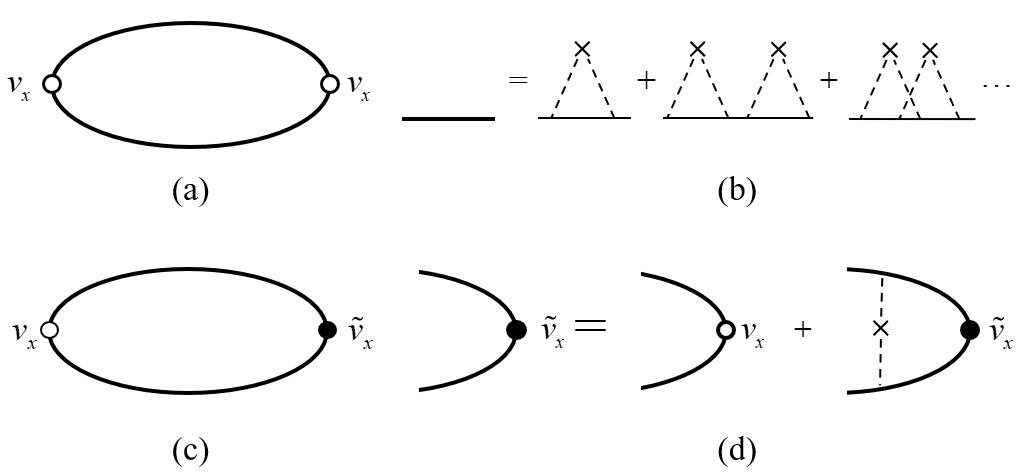}
		\caption{(a) The self-energy diagram (bubble diagram) of dc conductivity. (b) The self-energy modified Green's function, where the first wigwam diagram represents for the Born approximation. (c) The vertex corrected diagram of dc conductivity. (d) The ``dressed" vertex.}
		\label{fig:vertex-correction}
	\end{figure}
	The self-energy diagram (bubble diagram) of dc conductivity as shown in Fig. \ref{fig:vertex-correction}(a) can be calculated by
	\begin{equation}\begin{aligned}\label{Gaussian-dc-sigma-0}
			\sigma_{xx}^{0}(E)=\sigma_{xx}^{0,RA}(E)-{\rm Re}\left[\sigma_{xx}^{0,RR}(E)\right]
	\end{aligned}\end{equation}
	with
	\begin{eqnarray}
		\label{Gaussian-dc-sigma-RA}
		\sigma_{xx}^{0,RA}(E)&=&4\frac{e^2\hbar}{2\pi}\int\frac{d^2\bm{k}}{(2\pi)^2}{\rm Tr}\left[G^R(\bm{k},E)v_x(\bm{k})G^A(\bm{k},E)v_x(\bm{k})\right]_c\\
		\label{Gaussian-dc-sigma-RR}
		\sigma_{xx}^{0,RR}(E)&=&4\frac{e^2\hbar}{2\pi}\int\frac{d^2\bm{k}}{(2\pi)^2}{\rm Tr}\left[G^R(\bm{k},E)v_x(\bm{k})G^R(\bm{k},E)v_x(\bm{k})\right]_c
	\end{eqnarray}
	where the factor 4 denotes degeneracy of spin and valley, ${\rm Tr}[\cdots]$ means the trace in chiral basis, and the subscript $c$ indicates a disorder configuration average. We do the calculation in the eigen basis of pseudospin so that the velocity operator and Green's function are expressed as
	\begin{equation}\label{Gaussian-dc-vx}
		v_x(\bm{k})=v_f(\cos\theta_{\bm{k}}\sigma_z+\sin\theta_{\bm{k}}\sigma_y)=v_f\left(\begin{array}{cc}
			\cos\theta_{\bm{k}}&-i\sin\theta_{\bm{k}}\\i\sin\theta_{\bm{k}}&-\cos\theta_{\bm{k}}
		\end{array}\right)
	\end{equation}
	and
	\begin{equation}\label{Gaussian-dc-green}
		G^{L}(\bm{k},E)=\left(\begin{array}{cc}
			\frac{1}{E-E_{\bm{k}+}-\Sigma(\bm{k}+,E)}&0\\0&\frac{1}{E-E_{\bm{k}-}-\Sigma(\bm{k}-,E)}
		\end{array}\right)=\left(\begin{array}{cc}
			g^L_{+}(\bm{k},E)&0\\0&g^L_{-}(\bm{k},E)
		\end{array}\right)
	\end{equation}
	where $L=R,A$ denotes the retarded or advanced Green's function and $E_{\bm{k}\pm}=\pm\hbar v_fk$ is the eigenvalue. Plugging Eq.(\ref{Gaussian-dc-vx}) and Eq.(\ref{Gaussian-dc-green}) into Eq.(\ref{Gaussian-dc-sigma-RA}) and Eq.(\ref{Gaussian-dc-sigma-RR}), the value of $\sigma_{xx}^{0,RA}(\sigma_{xx}^{0,RR})$ is evaluated 
	\begin{equation}\begin{aligned}\label{Gaussian-dc-sigma-bubble}
			\sigma^{0,LM}_{xx}(E)=&\frac{2e^2\hbar v^2_f}{\pi}\int\frac{d^2\bm{k}}{(2\pi)^2}{\rm Tr}\left[
			\left(\begin{array}{cc}
				g^L_+&0\\0&g^L_-
			\end{array}\right)
			\left(\begin{array}{cc}
				\cos\theta_{\bm{k}}&-i\sin\theta_{\bm{k}}\\i\sin\theta_{\bm{k}}&-\cos\theta_{\bm{k}}
			\end{array}\right)
			\left(\begin{array}{cc}
				g^M_+&0\\0&g^M_-
			\end{array}\right)
			\left(\begin{array}{cc}
				\cos\theta_{\bm{k}}&-i\sin\theta_{\bm{k}}\\i\sin\theta_{\bm{k}}&-\cos\theta_{\bm{k}}
			\end{array}\right)
			\right]\\
			=&\frac{2e^2\hbar v^2_f}{\pi}\int\frac{d^2\bm{k}}{(2\pi)^2}\left[\cos^2\theta_{\bm{k}}(g^L_+g^M_++g^L_-g^M_-)+\sin^2\theta_{\bm{k}}(g^L_+g^M_-+g^L_-g^M_+)\right]\\
			=&\frac{e^2\hbar v^2_f}{\pi}\int\frac{kdk}{2\pi}\left[g^L_+g^M_++g^L_-g^M_-+g^L_+g^M_-+g^L_-g^M_+\right]\\
			=&\frac{e^2\hbar v^2_f}{\pi}\int\frac{kdk}{2\pi}(g^L_++g^L_-)(g^M_++g^M_-)\\
	\end{aligned}\end{equation}
	where $L,M=R,A$. Since we assume the self-energy function depending on both energy and  magnitude of wave vector as
	\begin{equation}\label{re-k-E}
		\Sigma(\bm{k}s,E)=\Sigma_1(E)-s\alpha\hbar v_fk+i\Sigma_2(E),
	\end{equation}
	the full Green's functions can be written as
	\begin{eqnarray}
		g^R_+(\bm{k},E)&=&\frac{1}{E-\hbar v_fk-{\rm Re}\Sigma(E,k+)-i{\rm Im}\Sigma(E,k+)}=\frac{1}{a-(1-\alpha)v_f\hbar k+i\eta}\\
		g^R_-(\bm{k},E)&=&\frac{1}{E+\hbar v_fk-{\rm Re}\Sigma(E,k-)-i{\rm Im}\Sigma(E,k-)}=\frac{1}{a+(1-\alpha)v_f\hbar k+i\eta}\\
		g^A_+(\bm{k},E)&=&\frac{1}{E-\hbar v_fk-{\rm Re}\Sigma(E,k+)+i{\rm Im}\Sigma(E,k+)}=\frac{1}{a-(1-\alpha)v_f\hbar k-i\eta}\\
		g^A_-(\bm{k},E)&=&\frac{1}{E+\hbar v_fk-{\rm Re}\Sigma(E,k-)+i{\rm Im}\Sigma(E,k-)}=\frac{1}{a+(1-\alpha)v_f\hbar k-i\eta}\\
	\end{eqnarray}
	with
	\begin{equation}
		\left\{\begin{array}{l}
			a=E-\Sigma_1(E)\\
			\eta=-\Sigma_2(E)\\
		\end{array}\right.
	\end{equation}
	
	Therefore we can obtain $\sigma^{0,RA}_{xx}(E)$ and ${\rm Re}\sigma^{0,RR}_{xx}(E)$ as
	\begin{equation}\begin{aligned}
			\sigma^{0,RA}_{xx}(E)=&\frac{e^2\hbar v^2_f}{\pi}\int\frac{kdk}{2\pi}\left[\frac{1}{a-(1-\alpha)\hbar v_fk+i\eta}+\frac{1}{a+(1-\alpha)\hbar v_fk+i\eta}\right]\\
			&\left[\frac{1}{a-(1-\alpha)\hbar v_fk-i\eta}+\frac{1}{a+(1-\alpha)\hbar v_fk-i\eta}\right]\\
			=&\frac{e^2\hbar v^2_f}{\pi}\int\frac{kdk}{2\pi}\frac{2(a+i\eta)}{(a+i\eta)^2-(1-\alpha)^2(\hbar v_fk)^2}\frac{2(a-i\eta)}{(a-i\eta)^2-(1-\alpha)^2(\hbar v_fk)^2}\\
			=&\frac{2e^2}{\pi\hbar}\frac{1}{(1-\alpha)^2}(a^2+\eta^2)\int^{\infty}_{\eta^2-a^2} dx\frac{1}{x^2+4a^2\eta^2}\\
			=&\frac{2e^2}{\pi h}\frac{1}{(1-\alpha)^2}(\frac{a}{\eta}+\frac{\eta}{a})\arctan\frac{a}{\eta}
	\end{aligned}\end{equation}
	and
	\begin{equation}\begin{aligned}
			{\rm Re}\sigma^{0,RR}_{xx}(E)=&{\rm Re}\frac{e^2\hbar v^2_f}{\pi}\int\frac{kdk}{2\pi}\left[\frac{1}{a-(1-\alpha)\hbar v_fk+i\eta}+\frac{1}{a+(1-\alpha)\hbar v_fk+i\eta}\right]^2\\
			=&{\rm Re}\frac{2e^2}{\pi h}\frac{1}{(1-\alpha)^2}\int dx\frac{(a+i\eta)^2}{\left[(a+i\eta)^2-x\right]^2}\\
			=&\frac{2e^2}{\pi h}\frac{1}{(1-\alpha)^2}\frac{(a+i\eta)^2}{(a+i\eta)^2-x}\bigg|^{\infty}_0\\
			=&-\frac{2e^2}{\pi h}\frac{1}{(1-\alpha)^2}
	\end{aligned}\end{equation}
	Thus the total bubble conductivity is 
	\begin{equation}
		\sigma^{0}_{xx}(E)=\sigma^{0,RA}_{xx}(E)-{\rm Re}[\sigma^{0,RR}_{xx}(E)]=\frac{2e^2}{\pi h}\frac{1}{(1-\alpha)^2}\left[1+(\frac{a}{\eta}+\frac{\eta}{a})\arctan\frac{a}{\eta}\right]
	\end{equation}
	from which we can obtain minimal conductivity at the limit $E\to0$
	\begin{equation}
		\sigma^{0}_{\text{min}}=\frac{2e^2}{\pi h}\frac{1}{(1-\alpha)^2}\left[1+\frac{\eta}{a}\frac{a}{\eta}\right]=\frac{4}{\pi}\frac{e^2}{h}\frac{1}{(1-\alpha)^2}
	\end{equation}
	
	The vertex correction of dc conductivity as shown in Fig. \ref{fig:vertex-correction}(c)(d), can be calculated by
	\begin{equation}\begin{aligned}\label{Gaussian-sigma-vertex}
			\sigma_{xx}^{v}(E)=\sigma_{xx}^{v,RA}(E)-{\rm Re}\left[\sigma_{xx}^{v,RR}(E)\right]
	\end{aligned}\end{equation}
	with
	\begin{eqnarray}
		\label{Gaussian-sigma-RA-vertex}
		\sigma_{xx}^{v,RA}(E)&=&4\frac{e^2\hbar}{2\pi}\int\frac{d^2\bm{k}}{(2\pi)^2}{\rm Tr}\left[G^R(\bm{k},E)v_x(\bm{k})G^A(\bm{k},E)\tilde{v}^{RA}_x(\bm{k})\right]\\
		\label{Gaussian-sigma-RR-vertex}
		\sigma_{xx}^{v,RR}(E)&=&4\frac{e^2\hbar}{2\pi}\int\frac{d^2\bm{k}}{(2\pi)^2}{\rm Tr}\left[G^R(\bm{k},E)v_x(\bm{k})G^R(\bm{k},E)\tilde{v}^{RR}_x(\bm{k})\right]
	\end{eqnarray}
	Here, the ``dressed" vertex function $\tilde{v}^{LM}_x$ is defined by self-consistent Bethe-Salpeter equation
	\begin{equation}\label{Gaussian-BS-equation}
		\tilde{v}^{LM}_x(\bm{k},E)=v_x(\bm{k})+\int\frac{d^2\bm{k}'}{(2\pi)^2}\mathcal{K}(\bm{k}-\bm{k}')U^{\dagger}_{\bm{k}}U_{\bm{k}'}G^L(\bm{k}',E)\tilde{v}^{LM}_x(\bm{k},E)G^M(\bm{k}',E)U^{\dagger}_{\bm{k}'}U_{\bm{k}}
	\end{equation}
	where $U^{\dagger}_{\bm{k}}U_{\bm{k}'}$ denotes the spin rotation while momentum changing
	\begin{equation}
		U_{\bm{k}}^{\dagger}U_{\bm{k}'}=
		\frac{1}{2}
		\left(\begin{array}{cc}
			1+e^{i\theta} & 1-e^{i\theta}\\1-e^{i\theta}&1+e^{i\theta}
		\end{array}\right)
	\end{equation}
	with $\theta=\theta_{\bm{k}'}-\theta_{\bm{k}}$. 
	In order to solve this self-consistent equation, we at first consider the first order approximation in the following
	\begin{equation}\begin{aligned}
			\tilde{v}^{(1),LM}_x(\bm{k},E)=&v_x(\bm{k})+\int\frac{d^2\bm{k}'}{(2\pi)^2}\mathcal{K}(\bm{k}-\bm{k}')U^{\dagger}_{\bm{k}}U_{\bm{k}'}G^L(\bm{k}',E)v_x(\bm{k})G^M(\bm{k}',E)U^{\dagger}_{\bm{k}'}U_{\bm{k}}\\
			=&v_x(\bm{k})+v_f\int\frac{d^2\bm{k}'}{(2\pi)^2}\frac{\mathcal{K}(\bm{k}-\bm{k}')}{4}
			\left(\begin{array}{cc}
				1+e^{i\theta} & 1-e^{i\theta}\\1-e^{i\theta}&1+e^{i\theta}
			\end{array}\right)
			\left(\begin{array}{cc}
				g^L_{+}&\\&g^L_{-}
			\end{array}\right)
			\left(\begin{array}{cc}
				\cos\theta_{\bm{k}'}&-i\sin\theta_{\bm{k}'}\\i\sin\theta_{\bm{k}'}&-\cos\theta_{\bm{k}'}
			\end{array}\right)\\
			&\left(\begin{array}{cc}
				g^M_{+}&\\&g^M_{-}
			\end{array}\right)
			\left(\begin{array}{cc}
				1+e^{-i\theta} & 1-e^{-i\theta}\\1-e^{-i\theta}&1+e^{-i\theta}
			\end{array}\right)\\
			=&v_x(\bm{k})+v_f\int\frac{d^2\bm{k}'}{(2\pi)^2}\frac{\mathcal{K}(\bm{k}-\bm{k}')}{4}
			\left(\begin{array}{cc}
				M_{11}&M_{12}\\M_{21}&M_{22}
			\end{array}\right)
	\end{aligned}\end{equation} 
	where the expressions of $M_{ss'}$ are in the following
	\begin{equation}\begin{aligned}
			M_{11}=&2\cos\theta(g^L_{+}g^M_{+}-g^L_{-}g^M_{-})\cos\theta_{\bm{k}}+\left[2\cos^2\theta(g^L_{+}g^M_{+}+g^L_{-}g^M_{-})+2\sin^2\theta(g^L_{+}g^M_-+g^L_{-}g^M_{+})\right]\cos\theta_{\bm{k}}\\
			M_{22}=&2\cos\theta(g^L_{+}g^M_{+}-g^L_{-}g^M_{-})\cos\theta_{\bm{k}}+\left[2\cos^2\theta(g^L_{+}g^M_{+}+g^L_{-}g^M_{-})+2\sin^2\theta(g^L_{+}g^M_-+g^L_{-}g^M_{+})\right](-\cos\theta_{\bm{k}})\\
			M_{12}=&2i\cos\theta(g^L_{+}g^M_--g^L_{-}g^M_{+})\sin\theta_{\bm{k}}+\left[2\cos^2\theta(g^L_{+}g^M_{+}+g^L_{-}g^M_{-})+2\sin^2\theta(g^L_{+}g^M_-+g^L_{-}g^M_{+})\right](-i\sin\theta_{\bm{k}})\\
			M_{21}=&2i\cos\theta(g^L_{+}g^M_--g^L_{-}g^M_{+})\sin\theta_{\bm{k}}+\left[2\cos^2\theta(g^L_{+}g^M_{+}+g^L_{-}g^M_{-})+2\sin^2\theta(g^L_{+}g^M_-+g^L_{-}g^M_{+})\right](i\sin\theta_{\bm{k}})
	\end{aligned}\end{equation}
	Based on the expressions of $M_{ss'}$, we can rewrite $\tilde{v}^{(1),LM}_x(\bm{k},E)$ as
	\begin{equation}\begin{aligned}\label{vertex-1}
			\frac{\tilde{v}^{(1),LM}_x(\bm{k},E)}{v_f}=&f^{LM}_0(\bm{k},E)\cos\theta_{\bm{k}}\sigma_0+f^{LM}_x(\bm{k},E)\sin\theta_{\bm{k}}\sigma_x+f^{LM}_y(\bm{k},E)\sin\theta_{\bm{k}}\sigma_y+f^{LM}_z(\bm{k},E)\cos\theta_{\bm{k}}\sigma_z
	\end{aligned}\end{equation}
	
	For $\tilde{v}^{(1),RA}_x(\bm{k},E)$, the functions $f^{RA}_i(\bm{k},E)$ with $(i=0,x,y,z)$ are
	\begin{equation}\begin{aligned}\label{f-s}
			f^{RA}_0(\bm{k},E)=&\int\frac{d^2\bm{k}'}{(2\pi)^2}\frac{\mathcal{K}(\bm{k}-\bm{k}')}{2}\cos\theta(g^R_{+}g^A_{+}-g^R_{-}g^A_{-})\\
			\approx&\frac{K_0(\hbar v_f)^2}{8\pi^2}\int k'dk'd\theta \left[1-\frac{\xi^2}{2}(k^2+k'^2)+\xi^2kk'\cos\theta\right]\cos\theta(g^R_{+}g^A_{+}-g^R_{-}g^A_{-})\\
			\approx&\frac{K_0\xi^2a}{4\pi(1-\alpha)^3\hbar v_f}\ln\left(\frac{E^2_c}{a^2+\eta^2}\right)k
	\end{aligned}\end{equation}
	\begin{equation}\begin{aligned}
			f^{RA}_x(\bm{k},E)=&\int\frac{d^2\bm{k}'}{(2\pi)^2}\frac{\mathcal{K}(\bm{k}-\bm{k}')}{2}i\cos\theta(g^R_{0,+}g^A_{0,-}-g^R_{0,-}g^A_{0,+})\\
			\approx&\frac{K_0(\hbar v_f)^2}{8\pi^2}\int k'dk'd\theta \left[1-\frac{\xi^2}{2}(k^2+k'^2)+\xi^2kk'\cos\theta\right]i\cos\theta(g^R_{+}g^A_{-}-g^R_{-}g^A_{+})\\
			\approx&\frac{K_0\xi^2\eta}{4\pi(1-\alpha)^3\hbar v_f}\ln\left(\frac{E^2_c}{a^2+\eta^2}\right)k
	\end{aligned}\end{equation}
	\begin{equation}\begin{aligned}
			f^{RA}_y(\bm{k},E)=f^{RA}_z(\bm{k},E)=&1+\int\frac{d^2\bm{k}'}{(2\pi)^2}\frac{\mathcal{K}(\bm{k}-\bm{k}')}{2}\left[\cos^2\Delta\theta(g^L_{+}g^M_{+}+g^L_{-}g^M_{-})+\sin^2\Delta\theta(g^L_{+}g^M_{-}+g^L_{-}g^M_{+})\right]\\
			\approx&1+\frac{K_0(\hbar v_f)^2}{8\pi}\int k'dk'(g^L_+g^M_{+}+g^L_{-}g^M_{-}+g^L_{+}g^M_{+}+g^L_{-}g^M_{-})\\
			\approx&1+\frac{K_0}{4\pi(1-\alpha)^2}(\frac{a}{\eta}+\frac{\eta}{a})\arctan\frac{a}{\eta}
	\end{aligned}\end{equation}
	
	For $\tilde{v}^{RR}_x(\bm{k},E)$, the functions $f^{RR}_i(\bm{k},E)$ with $(i=0,x,y,z)$ are
	\begin{equation}\begin{aligned}
			f^{RR}_0(\bm{k},E)=&\int\frac{d^2\bm{k}'}{(2\pi)^2}\frac{\mathcal{K}(\bm{k}-\bm{k}')}{2}\cos\theta(g^R_{+}g^R_{+}-g^R_{-}g^R_{-})\\
			\approx&\frac{K_0(\hbar v_f)^2}{8\pi^2}\int k'dk'd\theta \left[1-\frac{\xi^2}{2}(k^2+k'^2)+\xi^2kk'\cos\theta\right]\cos\theta(g^R_{+}g^R_{+}-g^R_{-}g^R_{-})\\
			\approx&\frac{K_0\xi^2(a+i\eta)}{4\pi(1-\alpha)^3\hbar v_f}\ln\left(\frac{E_c^2}{a^2+\eta^2}\right)k
	\end{aligned}\end{equation}
	\begin{equation}\begin{aligned}
			f^{RR}_x(\bm{k},E)=&\int\frac{d^2\bm{k}'}{(2\pi)^2}\frac{\mathcal{K}(\bm{k}-\bm{k}')}{2}i\cos\theta(g^R_{+}g^R_{-}-g^R_{-}g^R_{+})=0
	\end{aligned}\end{equation}
	
	\begin{equation}\begin{aligned}\label{f-f}
			f^{RR}_y(\bm{k},E)=f^{RR}_z(\bm{k},E)=&1+\int\frac{d^2\bm{k}'}{(2\pi)^2}\frac{\mathcal{K}(\bm{k}-\bm{k}')}{2}\left[\cos^2\theta(g^R_{+}g^R_{+}+g^R_{-}g^R_{-})+\sin^2\theta(g^R_{+}g^R_{-}+g^R_{-}g^R_{+})\right]\\
			\approx&1+\frac{K_0(\hbar v_f)^2}{8\pi}\int k'dk'(g^R_{+}g^R_{+}+g^R_{-}g^R_{-}+g^R_{+}g^R_{+}+g^R_{-}g^R_{-})\\
			\approx&1-\frac{K_0}{4\pi(1-\alpha)^2}
	\end{aligned}\end{equation}
	
	Plugging Eqs.(\ref{f-s})-(\ref{f-f}) into Eqs.(\ref{vertex-1}), (\ref{Gaussian-sigma-RA-vertex}), and (\ref{Gaussian-sigma-RR-vertex}), we can get the first-order $\sigma_{xx}^{v,RA}$ and ${\rm Re}(\sigma_{xx}^{v,RR})$ as
	\begin{equation}\begin{aligned}
			\sigma^{(1),RA}_{xx}(E)=&\frac{e^2\hbar v^2_f}{2\pi^2}\int kdk\left[1+\frac{K_0}{4\pi(1-\alpha)^2}(\frac{a}{\eta}+\frac{\eta}{a})\arctan\frac{a}{\eta}\right](g^R_{+}g^A_{+}+g^R_{-}g^A_{-}+g^R_{+}g^A_{-}+g^R_{-}g^A_{+})\\
			&+\frac{K_0\xi^2a}{4\pi(1-\alpha)^3\hbar v_f}\ln\left(\frac{E^2_c}{a^2+\eta^2}\right)k(g^R_{+}g^A_{+}-g^R_{-}g^A_{-})-i\frac{K_0\xi^2\eta}{4\pi(1-\alpha)^3\hbar v_f}\ln\left(\frac{E^2_c}{a^2+\eta^2}\right)k(g^R_{+}g^A_{-}-g^R_{-}g^A_{+})\\
			=&\frac{2e^2}{\pi h}\left\{(\frac{a}{\eta}+\frac{\eta}{a})\arctan\frac{a}{\eta}\left[1+\frac{K_0}{4\pi(1-\alpha)^4}(\frac{a}{\eta}+\frac{\eta}{a})\arctan\frac{a}{\eta}\right]+\frac{K_0\xi^2(a^2-\eta^2)}{4\pi(\hbar v_f)^2}\left[\frac{1}{(1-\alpha)^3}\ln\left(\frac{E^2_c}{a^2+\eta^2}\right)\right]^2\right\}
	\end{aligned}\end{equation}
	and
	\begin{equation}\begin{aligned}
			{\rm Re}[\sigma^{(1),RR}_{xx}(E)]=&\frac{e^2\hbar v^2_f}{2\pi^2}{\rm Re}\int kdk\left[1-\frac{K_0}{4\pi(1-\alpha)^2}\right](g^R_+g^R_++g^R_-g^R_-+g^R_+g^R_-+g^R_-g^R_+)\\
			&+\frac{K_0\xi^2(a+i\eta)}{4\pi(1-\alpha)^3\hbar v_f}\ln\left(\frac{E_c^2}{a^2+\eta^2}\right)k(g^R_+g^R_+-g^R_-g^R_-)\\
			=&\frac{2e^2}{\pi h}\left\{-1+\frac{K_0}{4\pi(1-\alpha)^4}+\frac{K_0\xi^2(a^2-\eta^2)}{4\pi(\hbar v_f)^2}\left[\frac{1}{(1-\alpha)^3}\ln\left(\frac{E^2_c}{a^2+\eta^2}\right)\right]^2\right\}
	\end{aligned}\end{equation}
	
	where $E_c=\hbar v_fk_c$. Thus the first-order vertex correction of dc conductivity is obtained as
	\begin{equation}\begin{aligned}
			\sigma_{xx}^{(1)}(E)=&\sigma_{xx}^{(1),RA}(E)-{\rm Re}\left[\sigma_{xx}^{(1),RR}(E)\right]\\
			=&\frac{2e^2}{\pi h}\left[1+(\frac{a}{\eta}+\frac{\eta}{a})\arctan\frac{a}{\eta}\right]+\frac{K_0}{4\pi(1-\alpha)^4}\left\{\left[(\frac{a}{\eta}+\frac{\eta}{a})\arctan\frac{a}{\eta}\right]^2-1\right\}
	\end{aligned}\end{equation}
	from which we can see that coefficients $f^{LM}_0$ and $f^{LM}_x$ can be ignored in the ``dressed" vertex. Thus, we can find that ``dressed" vertex has same matrix structure with the bare velocity and can be solved as
	\begin{equation}\begin{aligned}
			\tilde{v}^{RA}_x(\bm{k},E)=&\frac{1}{1-\frac{K_0}{4\pi(1-\alpha)^2}(\frac{a}{\eta}+\frac{\eta}{a})\arctan\frac{a}{\eta}}v_f(\cos\theta_{\bm{k}}\sigma_z+\sin\theta_{\bm{k}}\sigma_y)\\
			\tilde{v}^{RR}_x(\bm{k},E)=&\frac{1}{1+\frac{K_0}{4\pi(1-\alpha)^2}}v_f(\cos\theta_{\bm{k}}\sigma_z+\sin\theta_{\bm{k}}\sigma_y)
	\end{aligned}\end{equation}
	Based on this corrected vertex, the total dc conductivity will arrive at
	\begin{equation}
		\sigma^v_{xx}(E)=\frac{2e^2}{\pi h}\frac{1}{(1-\alpha)^2}\left[\frac{1}{1+\frac{K_0}{4\pi(1-\alpha)^2}}+\frac{(\frac{a}{\eta}+\frac{\eta}{a})\arctan\frac{a}{\eta}}{1-\frac{K_0}{4\pi(1-\alpha)^2}(\frac{a}{\eta}+\frac{\eta}{a})\arctan\frac{a}{\eta}}\right]
	\end{equation}
	and the corresponding vertex corrected minimal conductivity is
	%\begin{equation}
	%	\sigma^v_{\text{min}}=\frac{4e^2}{\pi h}\frac{1}{(1-\alpha)^2}\frac{1}{1-\left[\frac{K_0}{4\pi(1-\alpha)^2}\right]^2}
	%\end{equation}
	%which can revert to the result
	\begin{equation}
		\sigma^v_{\text{min}}=\frac{4e^2}{\pi h}\frac{1}{(1-\alpha)^2}\frac{1}{1-\left[\frac{K_0}{4\pi(1-\alpha)^2}\right]^2}\stackrel{K_0\lesssim1}{\longrightarrow}\frac{4e^2}{\pi h}\frac{1}{(1-\alpha)^2}
	\end{equation}
	Since typical experiment conditions of high mobility graphene correspond to $K_0\lesssim1$, the influence of vertex correction to the minimal conductivity is much smaller than those induced by the linear momentum dependent self-energy in the bubble diagram.
	\end{widetext}
	\end{appendix}

\end{document}